


\font\bigbold=cmbx12

\font\ninerm=cmr9
\font\eightrm=cmr8
\font\sixrm=cmr6
\font\fiverm=cmr5

\font\ninebf=cmbx9
\font\eightbf=cmbx8
\font\sixbf=cmbx6
\font\fivebf=cmbx5

\font\ninei=cmmi9  \skewchar\ninei='177
\font\eighti=cmmi8  \skewchar\eighti='177
\font\sixi=cmmi6    \skewchar\sixi='177
\font\fivei=cmmi5

\font\ninesy=cmsy9 \skewchar\ninesy='60
\font\eightsy=cmsy8 \skewchar\eightsy='60
\font\sixsy=cmsy6   \skewchar\sixsy='60
\font\fivesy=cmsy5

\font\nineit=cmti9
\font\eightit=cmti8

\font\ninesl=cmsl9
\font\eightsl=cmsl8

\font\ninett=cmtt9
\font\eighttt=cmtt8

\font\tenfrak=eufm10
\font\ninefrak=eufm9
\font\eightfrak=eufm8
\font\sevenfrak=eufm7
\font\fivefrak=eufm5

\font\tenbb=msbm10
\font\ninebb=msbm9
\font\eightbb=msbm8
\font\sevenbb=msbm7
\font\fivebb=msbm5

\font\tenssf=cmss10
\font\ninessf=cmss9
\font\eightssf=cmss8

\font\tensmc=cmcsc10


\newfam\bbfam
\textfont\bbfam=\tenbb
\scriptfont\bbfam=\sevenbb
\scriptscriptfont\bbfam=\fivebb
\def\Bbb{\fam\bbfam}

\newfam\frakfam
\textfont\frakfam=\tenfrak
\scriptfont\frakfam=\sevenfrak
\scriptscriptfont\frakfam=\fivefrak
\def\frak{\fam\frakfam}

\newfam\ssffam
\textfont\ssffam=\tenssf
\scriptfont\ssffam=\ninessf
\scriptscriptfont\ssffam=\eightssf
\def\ssf{\fam\ssffam}

\def\smc{\tensmc}


\def\eightpoint{%
\textfont0=\eightrm   \scriptfont0=\sixrm
\scriptscriptfont0=\fiverm  \def\rm{\fam0\eightrm}%
\textfont1=\eighti   \scriptfont1=\sixi
\scriptscriptfont1=\fivei  \def\oldstyle{\fam1\eighti}%
\textfont2=\eightsy   \scriptfont2=\sixsy
\scriptscriptfont2=\fivesy
\textfont\itfam=\eightit  \def\it{\fam\itfam\eightit}%
\textfont\slfam=\eightsl  \def\sl{\fam\slfam\eightsl}%
\textfont\ttfam=\eighttt  \def\tt{\fam\ttfam\eighttt}%
\textfont\frakfam=\eightfrak \def\frak{\fam\frakfam\eightfrak}%
\textfont\bbfam=\eightbb  \def\Bbb{\fam\bbfam\eightbb}%
\textfont\bffam=\eightbf   \scriptfont\bffam=\sixbf
\scriptscriptfont\bffam=\fivebf  \def\bf{\fam\bffam\eightbf}%
\abovedisplayskip=9pt plus 2pt minus 6pt
\belowdisplayskip=\abovedisplayskip
\abovedisplayshortskip=0pt plus 2pt
\belowdisplayshortskip=5pt plus2pt minus 3pt
\smallskipamount=2pt plus 1pt minus 1pt
\medskipamount=4pt plus 2pt minus 2pt
\bigskipamount=9pt plus4pt minus 4pt
\setbox\strutbox=\hbox{\vrule height 7pt depth 2pt width 0pt}%
\normalbaselineskip=9pt \normalbaselines
\rm}


\def\ninepoint{%
\textfont0=\ninerm   \scriptfont0=\sixrm
\scriptscriptfont0=\fiverm  \def\rm{\fam0\ninerm}%
\textfont1=\ninei   \scriptfont1=\sixi
\scriptscriptfont1=\fivei  \def\oldstyle{\fam1\ninei}%
\textfont2=\ninesy   \scriptfont2=\sixsy
\scriptscriptfont2=\fivesy
\textfont\itfam=\nineit  \def\it{\fam\itfam\nineit}%
\textfont\slfam=\ninesl  \def\sl{\fam\slfam\ninesl}%
\textfont\ttfam=\ninett  \def\tt{\fam\ttfam\ninett}%
\textfont\frakfam=\ninefrak \def\frak{\fam\frakfam\ninefrak}%
\textfont\bbfam=\ninebb  \def\Bbb{\fam\bbfam\ninebb}%
\textfont\bffam=\ninebf   \scriptfont\bffam=\sixbf
\scriptscriptfont\bffam=\fivebf  \def\bf{\fam\bffam\ninebf}%
\abovedisplayskip=10pt plus 2pt minus 6pt
\belowdisplayskip=\abovedisplayskip
\abovedisplayshortskip=0pt plus 2pt
\belowdisplayshortskip=5pt plus2pt minus 3pt
\smallskipamount=2pt plus 1pt minus 1pt
\medskipamount=4pt plus 2pt minus 2pt
\bigskipamount=10pt plus4pt minus 4pt
\setbox\strutbox=\hbox{\vrule height 7pt depth 2pt width 0pt}%
\normalbaselineskip=10pt \normalbaselines
\rm}


\def\pagewidth#1{\hsize= #1}
\def\pageheight#1{\vsize= #1}
\def\hcorrection#1{\advance\hoffset by #1}
\def\vcorrection#1{\advance\voffset by #1}

\newif\iftitlepage   \titlepagetrue               
\newtoks\titlepagefoot     \titlepagefoot={\hfil} 
\newtoks\otherpagesfoot    \otherpagesfoot={\hfil\tenrm\folio\hfil}
\footline={\iftitlepage\the\titlepagefoot\global\titlepagefalse
           \else\the\otherpagesfoot\fi}

\font\extra=cmss10 scaled \magstep0
\setbox1 = \hbox{{{\extra R}}}
\setbox2 = \hbox{{{\extra I}}}
\setbox3 = \hbox{{{\extra C}}}
\setbox4 = \hbox{{{\extra Z}}}
\setbox5 = \hbox{{{\extra N}}}

\def\RRR{{{\extra R}}\hskip-\wd1\hskip2.0
   true pt{{\extra I}}\hskip-\wd2
\hskip-2.0 true pt\hskip\wd1}
\def\Real{\hbox{{\extra\RRR}}}    

\def\CCC{{{\extra C}}\hskip-\wd3\hskip 2.5 true pt{{\extra I}}
\hskip-\wd2\hskip-2.5 true pt\hskip\wd3}
\def\Complex{\hbox{{\extra\CCC}}\!\!}   




\def\R{{\Real}}
\def\C{{\Complex}}

\def\ve{\vfill\eject}

\def\frac#1#2{{#1\over#2}}

\def\({\left(}
\def\){\right)}

\def\pmb#1{\setbox0=\hbox{$#1$}%
   \kern-.025em\copy0\kern-\wd0
   \kern.05em\copy0\kern-\wd0
   \kern-.025em\raise.0433em\box0 }


\def\abstract#1{
{\parindent=30pt\narrower\noindent\ninepoint\openup
2pt #1\par}}


\newcount\notenumber\notenumber=1
\def\note#1
{\unskip\footnote{$^{\the\notenumber}$}
{\eightpoint\openup 1pt #1}
\global\advance\notenumber by 1}


\global\newcount\secno \global\secno=0
\global\newcount\meqno \global\meqno=1
\global\newcount\appno \global\appno=0
\newwrite\eqmac
\def\romappno{\ifcase\appno\or A\or
B\or C\or D\or E\or F\or G\or H
\or I\or J\or K\or L\or M\or N\or
O\or P\or Q\or R\or S\or T\or U\or
V\or W\or X\or Y\or Z\fi}
\def\eqn#1{
        \ifnum\secno>0
            \eqno(\the\secno.\the\meqno)
\xdef#1{\the\secno.\the\meqno}
          \else\ifnum\appno>0
            \eqno(
{\rm\romappno}.\the\meqno)\xdef#1{{\rm\romappno}.\the\meqno}
          \else
            \eqno(\the\meqno)\xdef#1{\the\meqno}
          \fi
        \fi
\global\advance\meqno by1 }


\global\newcount\refno
\global\refno=1 \newwrite\reffile
\newwrite\refmac
\newlinechar=`\^^J
\def\ref#1#2{\the\refno\nref#1{#2}}
\def\nref#1#2{\xdef#1{\the\refno}
\ifnum\refno=1\immediate\openout\reffile=refs.tmp\fi
\immediate\write\reffile{
     \noexpand\item{[\noexpand#1]\ }#2\noexpand\nobreak.}
     \immediate\write\refmac{\def\noexpand#1{\the\refno}}
   \global\advance\refno by1}
\def\semi{;\hfil\noexpand\break ^^J}
\def\nl{\hfil\noexpand\break ^^J}
\def\refn#1#2{\nref#1{#2}}

\def\jpA#1#2#3{{\it J.  Phys.} {\bf A{#1}} ({#2}) #3}
\def\ijmp#1#2#3{{\it Int.  J.  Mod.  Phys.} {\bf A{#1}} ({#2}) #3}

\def\np#1#2#3{{\it Nucl.  Phys.} {\bf B{#1}} ({#2}) #3}

\def\plA#1#2#3{{\it Phys.  Lett.} {\bf {#1}A} ({#2}) #3}

\def\prl#1#2#3{{\it Phys.  Rev.  Lett.} {\bf #1} ({#2}) #3}


{
\refn\RS
{M. Reed and B. Simon,
\lq\lq Methods of Modern Mathematical Physics\rq\rq, Vol.I, II,
Academic Press, New York, 1980}

\refn\AGHH
{S. Albeverio, F. Gesztesy, R. H{\o}egh-Krohn and H. Holden,
\lq\lq Solvable Models in Quantum Mechanics\rq\rq,
Springer, New York, 1988}

\refn\AK
{S. Albeverio and P. Kurasov,
\lq\lq
Singular Perturbations of Differential Operators
\rq\rq,
Cambridge Univ. Press,
Cambridge, 2000}

\refn\AEL
{J.E. Avron, P. Exner and Y. Last,
\prl{72}{1994}{869}}

\refn\KI
{A. Kiselev, J. Math. Anal. Appl. {\bf 212}
(1997) 263.}

\refn\CHughes
{P.R. Chernoff and R.J. Hughes,
{\it J. Funct. Anal.} {\bf 111} (1993) 97}

\refn\CSa
{T. Cheon and T. Shigehara,
\plA{243}{1998}{111}}

\refn\AN
{S. Albeverio and L. Nizhnik,
{\sl Approximation of general Zero-Range Potentials},
Uni. Bonn Preprint no.585 (1999)}

\refn\Jackiwb
{R. Jackiw,
in \lq\lq Diverse topics in
Theoretical and Mathematical Physics\rq\rq,
World Scientific, Singapore, 1995}

\refn\CSb
{T. Cheon and T. Shigehara,
\prl{82}{1999}{2536}}

\refn\CHa
{T. Cheon,
\plA{248}{1998}{285}}

\refn\TFC
{I. Tsutsui, T. F\"{u}l\"{o}p and T. Cheon,
{\it Duality and Anholonomy
in Quantum Mechanics
of 1D
Contact Interactions}, KEK Preprint 2000-3,
quant-ph/0003069
}

\refn\ADK
{S. Albeverio, L. Dabrowski and P. Kurasov,
{\it Lett. Math. Phys.} {\bf 45} (1998) 33}

\refn\FT
{T. F\"{u}l\"{o}p and I. Tsutsui,
\plA{264}{2000}{366}}

\refn\Seba
{P. \v{S}eba,
{\it Czech. J. Phys.} {\bf 36} (1986) 667}

\refn\ABD
{S. Albeverio, Z. Brze\'{z}niak and L. Dabrowski,
{\it J. Funct. Anal.} {\bf 130} (1995) 220}

\refn\RT
{J.M. Rom\'{a}n and R. Tarrach,
\jpA{29}{1996}{6073}}

\refn\FG
{E. Farhi and S. Gutmann,
\ijmp{5}{1990}{3029}}

\refn\SW
{A. Shapere and F. Wilczek,
\lq\lq Geometric Phases in Physics\rq\rq,
World Scientific, Singapore, 1989}

\refn\EG
{P. Exner and H. Grosse,
{\it Some properties of the one-dimensional
generalized point
interactions (a torso)},
math-ph/9910029
}

\refn\Junker
{G. Junker,
\lq\lq Supersymmetric Methods in Quantum
and Statistical
Physics\rq\rq,
Texts and Monographs in Physics, Springer,
Berlin, 1996}

\refn\Wit
{E. Witten,
\np{188}{1988}{513}}

\refn\BBBJR
{A. Bertoni, P.Bordone, R.Brunetti, C.Jacoboni and
S.Reggiani,
\prl{84}{2000}{5912}}

}


\def\mod{\hbox{mod }}

\def\Re{\hbox{Re}}
\def\n{_{new}}
\def\<{\langle}
\def\>{\rangle}
\def\=>{\Rightarrow}
\def\==>{\Longrightarrow}
\def\++{^{(+)}}
\def\--{^{(-)}}

\def\[{\left[}
\def\]{\right]}
\def\ddef#1#2#3#4{\left\{ \eqalign{ #1 & \qquad #2 \cr %
    #3 & \qquad #4 } \right.}


\def\today{\space\number\day\
\ifcase\month\or
January\or February\or March\or April\or
May\or June\or July\or August\or September\or
October\or November\or December\fi\
\number\year}



\pageheight{23cm}
\pagewidth{14.8cm}
\hcorrection{0mm}
\magnification= \magstep1
\def\bsk{%
\baselineskip= 16.8pt plus 1pt minus 1pt}
\parskip=5pt plus 1pt minus 1pt
\tolerance 6000


\null

{
\leftskip=100mm
\hfill\break
KEK Preprint 2000-54
\hfill\break
\par}
{\baselineskip=18pt

\centerline{\bigbold
Symmetry, Duality and Anholonomy}
\centerline{\bigbold
of Point Interactions in One Dimension
}

\vskip 10pt

\centerline{
\smc
Taksu Cheon\note
{email:\quad cheon@mech.kochi-tech.ac.jp}
}

\vskip 3pt
{
\baselineskip=13pt
\centerline{\it Laboratory of Physics}
\centerline{\it Kochi University of Technology}
\centerline{\it Tosa Yamada, Kochi 782-8502, Japan}
}

\vskip 3pt

\centerline{
\smc
Tam\'{a}s F\"{u}l\"{o}p\note
{email:\quad fulopt@poe.elte.hu}
}

\vskip 3pt
{
\baselineskip=13pt
\centerline{\it Institute for Theoretical Physics}
\centerline{\it Roland E\"{o}tv\"{o}s University}
\centerline{\it H-1117 Budapest, P\'{a}zm\'{a}ny
P. s\'{e}t\'{a}ny 1/A, Hungary}
}

\vskip 3pt
\centerline{\rm and}
\vskip 3pt

\centerline{
\smc
Izumi Tsutsui\note
{email:\quad izumi.tsutsui@kek.jp}
}

\vskip 5pt

{
\baselineskip=13pt
\centerline{\it
Institute of Particle and Nuclear Studies}
\centerline{\it
High Energy Accelerator Research Organization (KEK)}
\centerline{\it Tsukuba 305-0801, Japan}
}

\vskip 15pt

\abstract{%
{\bf Abstract.}\quad
We analyze the spectral structure of the
one dimensional quantum mechanical system with
point interaction,
which is known to be parametrized
by the group $U(2)$.
Based on the classification of the interactions
in terms of symmetries,
we show, on a general ground, how the fermion-boson
duality and the spectral anholonomy recently 
discovered can arise.
A vital role is played by a hidden $su(2)$ formed by
a certain set of discrete transformations,
which becomes a symmetry if the point 
interaction belongs to
a distinguished $U(1)$ subfamily in 
which all states are doubly
degenerate.  Within the $U(1)$, 
there is a particular interaction
which admits the interpretation of 
the system as a supersymmetric
Witten model.
}

\vskip 10pt
{\ninepoint
\indent{PACS codes: 3.65.-w, 2.20.-a, 73.20.Dx\hfill\break}
\indent{Keywords: Point interaction, Duality, Anholonomy}
}
}


\pageheight{23cm}
\pagewidth{15.7cm}
\hcorrection{-1mm}
\magnification= \magstep1
\def\bsk{%
\baselineskip= 16.4pt plus 1pt minus 1pt}
\parskip=5pt plus 1pt minus 1pt
\tolerance 8000
\bsk


\ve
\secno=1 \meqno=1


\centerline
{\bf 1. Introduction}
\medskip

The point interaction, {\it i.e.}, the interaction
of zero range, is perhaps the simplest
among all interactions in physics, and yet it is
the most generic in the sense that any local potential
can be approximated by the point interaction in the long
wavelength limit.
The formulation of the point interaction
in quantum theory requires
some sort of treatment of the singularity that
appears in the short range limit of the potential.
The problem of short range singularities has been
familiar in quantum
field theory where the ultraviolet divergence
is tamed by the procedure of renormalization.

It is not widely recognized, however,
that the same delicate problem
already exists in the treatment of a quantum
mechanical particle under point (or contact) interaction.
In spatial dimension one, a stable zero range limit
exists for the family of finite
range interactions with a constant volume integral of
the strength, which is
none other than the Dirac $\delta$-function interaction.
However, its straightforward extension to spacial dimensions
two and three fails because of the
divergence in the Green's function.
To put the point interaction on a
mathematically sound basis, and thereby uncover its
entire physical content
in arbitrary dimensions,
one has to resort to a more rigorous approach
based on functional analysis.
With the theory of self-adjoint extensions [\RS],
one finds that the quantum mechanical
point interaction can be defined in dimension one, 
two and three,
and it is nontrivial in each of these cases.
In particular,
in one dimension there arises a $U(2)$ family of
interactions [\AGHH] (see also [\AK]), where besides the
Dirac $\delta$-function interaction which
induces
discontinuity in the derivative of the wave function,
there is another type of
point interaction, called
`$\varepsilon$-function interaction',
which induces discontinuity
in the wave function itself.
{}From these two typical zero-range forces, one can construct
any of the interactions in the $U(2)$ family
which in general
has discontinuity both in the wave function
and its derivative.  Although these generic
point interactions predict
several unusual properties which
defy our intuition on pointlike objects [\AK,\AEL,\KI],
the fact that these interactions are fully
expressible as a limit of local
potentials [\CHughes,\CSa,\AN] suggests that
their experimental realization is feasible, especially
in view of the rapid advance of nano-scale technology
of recent years.

It has been known
that, in dimension two and three where the family of
point interactions is given by $U(1)$,
the realization of the interactions
can be understood through
coupling constant renormalization
analogous to that encountered
in quantum field theory, and that in two dimensions
this offers a prototype of
quantum anomaly by the breakdown
of scale symmetry [\Jackiwb].  In dimension one,
on the other hand, the larger family
$U(2)$ of allowed interactions
admits a number of more intriguing
features.
These include
the \lq fermion-boson duality', which is the
phenomenon that two systems with distinct point
interactions related by coupling inversion
shares an identical spectrum with
symmetric and antisymmetric states interchanged [\CSb].
The other notable feature is
the spectral anholonomy [\CHa], which is the appearance
of a double spiral structure of the energy levels
when the subfamily of parity
invariant point interactions is considered.  However,
so far these
phenomena have been shown to arise for a specific set
of interactions, and one may wonder if this
is a generic aspect of the point interaction
rather than something accidental.

The aim of the present paper is to provide a coherent
framework to understand all of these features on
a general basis extending our previous report [\TFC]
substantially, and thereby show that these are in fact
a generic aspect of the point interaction in one dimension.
Since these features are intimately related to
symmetries of the system, we shall
first study
symmetries and
associated invariant subfamilies of the point interactions.
The symmetries with respect to
parity, time reversal and Weyl scaling transformations
have been discussed earlier
in [\ADK] using a local description of the family.
Here we give a fuller account of them by adopting a
description valid globally over the $U(2)$ family, and
point out that behind the aforementioned
nontrivial features underlies an $su(2)$ structure
consisting of generators of
two novel discrete transformations along with
parity.  The $su(2)$ will then be shown to
become a symmetry if the point interactions
belong to a distinct $U(1)$
subfamily in the $U(2)$.  The $U(1)$
subfamily forms a singular
circle of spectral degeneracy
in the parameter space, which is crucial in realizing
the duality and the anholonomy.
We will also touch upon briefly
the possibility
of supersymmetry, which
is suggested from the degeneracy
in the $U(1)$ subfamily.

The plan of the paper is as follows.
After this Introduction, we present in Sect.2
a concise account of how the $U(2)$ family
of point interactions appears in one dimension.
Our argument is elementary [\FT] and based on the unitarity
(probability conservation) of the system, which is,
in essence, equivalent to the requirement that
the Hamiltonian be self-adjoint.  The gap between
our derivation and the standard,
more involved one which employs
the theory of self-adjoint extensions will be
filled by Appendix A supplied at the end of the paper.
In Sect.3, we discuss symmetries of the
point interactions in detail.  The crucial ingredient
here is the existence of the $su(2)$ generators
which give rise to spectrum-preserving transformations
in the $U(2)$ family.
We first examine the spectral characteristics
such as the spectral flows
of the invariant subfamily
under the parity transformation.
We then move on to the generic
invariant subfamily
associated with the $su(2)$ generators.
This will be important in Sect.4 when we discuss
the duality, which can be
seen in any of the subfamilies in the $su(2)$.
The parity invariant
subfamily is again considered in analyzing
the spectral anholonomy in the double spiral form,
while the Weyl scaling invariant subfamily
is found to be
the parameter subspace where the Berry phase
anholonomy can be found.
Furthermore, supersymmetry
is observed for a particular point interaction
belonging to the $U(1)$ subfamily where the states
are all classified according to the \lq spin' of
the $su(2)$ symmetry.
For illustration of these phenomena we sometimes
put the particle
in a box to obtain an entirely discrete spectrum.
The necessary material for the spectrum as well as
other basic physical properties of the system
occurring under the point interaction is provided
in Appendix B.
Finally, Sect.5
is devoted to our Summary and Discussions.

\ve
\secno=2 \meqno=1


\centerline
{\bf 2. Systems with Point Interaction}
\medskip

We begin by reviewing how to describe
and characterize one-dimensional point interactions
in quantum mechanics.  The requirement of
probability conservation is shown to determine
the allowed class of point interactions
to be the group $U(2)$.
The local description
of point interactions often found
in the literature is then related to
our global description given by the $U(2)$.
The set of point
interactions missing in the local description
is shown to form a subgroup
$U(1) \times U(1) \subset U(2)$.

\bigskip
\noindent
{\bf 2.1. Point interaction in quantum mechanics}
\medskip

Given a quantum system with point interaction in one
dimension,
the first question we
encounter is how to describe and characterize
the interaction in an inclusive and consistent manner.
In other words,
we want to know what kind of
point interactions are allowed
quantum mechanically on a line, and how to specify them
mathematically.

To answer this, we first need to clarify what we mean
by `point interaction'.  As discussed in the
Introduction, we regard it
to be an idealized long
wavelength or infrared limit of localized
interactions
in one dimension,
and hence it is a singular interaction with zero
range occurring at one point, say $x = 0$, on a line
$\R$.  A system with such an interaction
can be
described by a free system on the line with the
singular point removed, namely, on
$\R\setminus\{0\}$.  Once defined this way, our main
concern will be the Hamiltonian operator,
$$
H =
-{{\hbar^2}\over{2m}}{{d^2}\over{dx^2}} \ ,
\eqn\ham
$$
defined on a proper domain $D(H)$ in the
Hilbert space,
$$
{\cal H} = L^2(\R\setminus\{0\})\ .
\eqn\hsp
$$
The Hamiltonian $H$ is meant to be an
observable of our quantum system, and if this
is the sole requirement, then
our question boils down to
the problem of seeking
for the allowed class of domains $D(H)$
on which the Hamiltonian operator (\ham)
becomes self-adjoint.
In physical terms, being the generator
for time evolution,
a self-adjoint Hamiltonian implies probability
conservation in the entire system.
Probability
conservation is guaranteed if
the probability current,
$$
j(x) = - {{i\hbar}\over{2m}}\left(
(\varphi^*)'\varphi - \varphi^* \varphi'
\right)(x)\ ,
\eqn\ctwall
$$
is continuous around the singular
point, namely,
$$
j(0_-) = j(0_+)\ ,
\eqn\pccond
$$
where $0_+$ and $0_-$ denote the limits to zero
from above and from below, respectively.
\topinsert
\vskip 1.5cm
\let\picnaturalsize=N
\def\picsize{4cm}
\def\picfilename{f1.epsf}
\input epsf
\ifx\picnaturalsize N\epsfxsize \picsize\fi
\hskip 2.8cm\epsfbox{\picfilename}
\vskip 0.5cm
\abstract{%
{\bf Figure 1.}
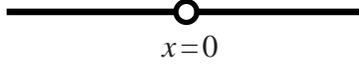
Point interaction in one dimension.  The interaction
that occurs only at $x = 0$ may be realized simply by
removing the point from
the real line $\R$.  The unitarity of the system
demands that the probability current $j(x)$ be continuous
at the missing point.
}
\endinsert

The requirement (\pccond) implies that
any state in the domain
$D(H)$ must obey a certain set
of  boundary conditions at $x = 0_\pm$.
To exhibit those
conditions explicitly, we
follow the approach of Ref.[\FT] and
define the two-component vectors,
$$
\Phi :=
  \left( {\matrix{{\varphi (0_+)}\cr
                  {\varphi (0_-)}\cr}
         }
  \right),
\qquad
\Phi' :=
  \left( {\matrix{{ \varphi' (0_+)}\cr
                  {-\varphi' (0_-)}\cr}
         }
  \right) ,
\eqn\vectors
$$
to a given state $\varphi$.
(Note the minus sign in the second component of $\Phi'$.)
In terms of these vectors,
the requirement (\pccond) becomes
$
\, \Phi'^\dagger \Phi = \Phi^\dagger \Phi' \, ,
$
which is equivalent to
$$
|\Phi-i L_0 \Phi'| = |\Phi+i L_0 \Phi'|\ ,
\eqn\samemodulo
$$
with $L_0$ being an arbitrary nonzero constant with
the dimension of length.
This means that, with a two-by-two unitary matrix
$U\in U(2)$, we have
$$
(U-I)\Phi+iL_0(U+I)\Phi'=0\ .
\eqn\unitrel
$$
It can be shown (see the Appendix)
that this matrix $U$ is the same for
all states $\varphi$ in the domain
$D(H)$.
A standard
parametrization for $U \in U(2)$ is given by
$$
U= e^{i \xi} \pmatrix{ \alpha  & \beta    \cr
                           -\beta^* & \alpha^* \cr}
= e^{i\xi } \left( {\matrix
            {{ \alpha_R+i\alpha_I}&{\beta_R+i\beta_I}\cr
             {-\beta_R+i\beta_I}&{\alpha_R-i\alpha_I}\cr}
         }
  \right)
\ ,
\eqn\stparm
$$
where $\xi \in [0,\pi)$ and
$\alpha$,
$\beta$ are complex
parameters satisfying
$$
\vert \alpha\vert^2 + \vert \beta\vert^2 =
\alpha_R^2+\alpha_I^2+\beta_R^2+\beta_I^2 = 1\ .
\eqn\parcon
$$
Since there is a one-to-one
correspondence between a physically distinct point
interaction and a self-adjoint Hamiltonian,
the foregoing argument
shows that the entire family
$\Omega$ of point interactions admitted in quantum
mechanics is exhausted by the group $U(2)$,
and that
different point interactions are characterized
by different boundary conditions (\unitrel)
for the wave functions.  In short,
in quantum mechanics
a point interaction in one dimension
is specified by
its {\it characteristic matrix} $U \in U(2)$.
(Alternatively, the condition (\unitrel) can
be derived using the theory
of self-adjoint extensions [\AGHH].)

In passing
we point out that the parameter $L_0$ shows the presence
of a scale factor in the system and, consequently,
signals the quantum mechanical
breakdown of scaling symmetry that is
present on the classical level.
Scale invariance remains valid quantum
mechanically only for a certain
subclass of point interactions (for
detail, see Sect.3).  The parameter $L_0$
adds no freedom to the boundary
condition other than the $U(2)$, because
its apparent freedom can be shown to be
absorbed by adjusting the $U(2)$ parameters
in $U$ (see the Appendix).

\bigskip
\noindent
{\bf 2.2. Global vs. local description}
\medskip

In the literature [\CHughes,\ADK,\Seba,\ABD,\RT],
a different parametrization for the $U(2)$ point interactions
has often been employed.  It is the description
in terms of the connection condition
at the singular point,
$$
  \left( {\matrix{{\varphi (0_+)}\cr
                  {\varphi' (0_+)}\cr}
         }
  \right)
=
  \Lambda \left( {\matrix{{ \varphi (0_-)}\cr
                  {\varphi' (0_-)}\cr}
         }
  \right)\ .
\eqn\trcon
$$
The (infinitesimal) transfer matrix $\Lambda$
used in (\trcon) takes the form
$$
\Lambda = e^{i\chi}
\pmatrix{ a & b \cr
          c & d \cr}\ , \qquad \chi \in [0, \pi),
\quad
a, b, c, d \in \R, \quad ad - bc = 1\ ,
\eqn\matleq
$$
which shows that the family
of point interactions covered by (\trcon)
forms the group $U(1) \times SL(2, \R)$ rather than
$U(2)$.
Actually, it is straightforward to confirm
that,
for $\beta \ne 0$, our description
(\unitrel) with (\stparm) can be put into
the connection form (\trcon) with
$$
\Lambda = {i\over{\beta_R -i\beta_I}}
\pmatrix{ \sin\xi - \alpha_{I}
                & -L_0(\cos\xi + \alpha_{R}) \cr
           L_0^{-1}(\cos\xi - \alpha_{R})
                &  \sin\xi + \alpha_{I} \cr}\ .
\eqn\matl
$$

The transfer matrix provides a direct description
of the physical effect that arises
at the singular point,
and for this reason the description (\trcon)
is convenient to use for
characterizing the point interaction based on the
scattering (reflection and transmission)
picture of incident waves (see Appendix B).
However, it should be stressed that
the transfer matrix provides a
{\it local} description of
the entire family $\Omega \simeq U(2)$ of
point interactions,
since
it does not contain the case $\beta = 0$
as seen from the above correspondence.  To find out
what kind of point interactions are missing
from (\trcon), we restrict ourselves to the case
$\beta = 0$ where we can put
$\alpha_R+i\alpha_I
= e^{i\rho}$ with $\rho \in [0, 2\pi)$
to obtain
$U = e^{i\xi}\, e^{i\rho\sigma_3}$ in terms of
the Pauli matrix $\sigma_3$.  We will also
need
the projection matrices with respect to
$\sigma_i$,
$$
P^\pm_i := {{1 \pm\sigma_i}\over 2}\ ,
\qquad \hbox{for} \quad i = 1, \, 2, \, 3,
\eqn\prjthree
$$
each of which satisfies
$$
(P^\pm_i)^2 = P^\pm_i\ ,
\qquad P^\pm_i P^\mp_i = 0\ , \qquad
P^+_i + P^-_i = 1\ .
\eqn\propr
$$

A convenient parametrization
available for the present case is then
provided by using
$ \phi_\pm = \xi \pm \rho \; (\mod 2 \pi) $ to express
$$
U = U(\phi_+, \phi_-) =
e^{i(\phi_+P^+_3 + \phi_-P^-_3)}\ .
\eqn\usep
$$
Note that the ranges $\phi_\pm \in [0, 2\pi)$
cover the entire $U$ belonging to
this case, for which
we have the chiral decomposition,
$$
U(\phi_+, \phi_-)
= U_+(\phi_+)\, U_-(\phi_-)
= U_-(\phi_-)\, U_+(\phi_+)\ ,
\eqn\chdecp
$$
with
$$
U_\pm(\phi_\pm) := e^{i\phi_\pm P_3^\pm}
=  1 + (e^{i\phi_\pm} - 1) P_3^\pm\ .
\eqn\prchr
$$
This shows that the set of
these $U$ forms a subfamily given by the
subgroup,
$$
\Omega_R \simeq U(1) \times U(1) \subset \Omega
\simeq U(2)\ .
\eqn\subthree
$$
We shall find in the next section that
$\Omega_R$ appears also as
the subfamily of point interactions invariant
under a certain discrete transformation.  These
point interactions are those of a perfect wall
placed at $x = 0$
through which no probability flow is allowed,
separating the left $\R^-$
and the right half line $\R^+$ completely.

\ve
\secno=3 \meqno=1


\centerline
{\bf 3. Symmetries}
\medskip

In this section we study
how symmetry transformations such as
parity, time-reversal and Weyl scaling
are implemented for point interactions.
Our analysis is in a sense analogous to
Ref.[\ADK], but we shall soon see the advantage
of the global nature of our parametrization.
Subfamilies invariant under
these transformations are defined, which
are then used to classify the point interactions
in one dimension.
Behind these subfamilies is an
$su(2)$ structure, which plays a central
role in establishing the duality to be
discussed later.  The structure of the
invariant subfamilies obtained here
will also be analyzed.

\bigskip
\noindent
{\bf 3.1. Transformations and invariant subfamilies}
\medskip

\noindent
{\it 3.1.1. Parity}

To begin with, we consider the
{\it parity} transformation which is defined
for any $\varphi \in {\cal H}$ by
$$
{\cal P}: \quad
\varphi(x) \longrightarrow
({\cal P}\varphi)(x) := \varphi( - x)\ .
\eqn\parity
$$
Under this, the boundary values of a wave function
and its derivatives
at $x = 0_\pm$ are interchanged.  Accordingly, the
boundary vectors (\vectors) transform as
$$
\Phi
\buildrel {\cal P} \over \longrightarrow \sigma_1 \Phi\ ,
\qquad
\Phi'
\buildrel {\cal P} \over \longrightarrow \sigma_1 \Phi'\ .
\eqn\vecpty
$$
As can be seen from (\unitrel),
this induces a map on the $U(2)$ family $\Omega$
such that
$$
U \buildrel {\cal P} \over \longrightarrow
\sigma_1\, U\, \sigma_1\ ,
\eqn\parityonu
$$
that is,
a point interaction (or a domain
$D(H)$) specified by $U$ maps to another one specified by
$\sigma_1 U \sigma_1$. {}From this it follows that
parity invariant
point interactions (or parity invariant domains $D(H)$)
are characterized by those $U$ which fulfill
$$
\sigma_1\, U\, \sigma_1 = U\ .
\eqn\pinv
$$
The general solution of this is given by
$\alpha_I = 0$ and $\beta_R = 0$ in
(\stparm), and if we put $\alpha_R = \cos\phi$
and $\beta_{I} = \sin\phi$, we have
$$
U = e^{i\xi}
\pmatrix{ \cos\phi    & i\sin\phi \cr
          i\sin\phi  & \cos\phi  \cr} \ ,
\eqn\tsol
$$
which can be cast into a form analogous to
(\usep) as
$$
U = U(\theta_+, \theta_-)
  = e^{i(\theta_+P^+_1 + \theta_-P^-_1)} \ ,
\eqn\chiralproj
$$
using angle parameters
$\theta_\pm \in [0, 2\pi)$ defined by
$$
\theta_\pm := \xi \pm \phi,
\eqn\no
$$
and the
projection matrices,
$P^\pm_1$ in (\prjthree).
Thus, by an analogous reasoning, we observe that
the {\it parity invariant} subfamily
$\Omega_{ P}$ is a direct product
of the two $U(1)$ groups,
$$
\Omega_{ P} \simeq U(1) \times U(1)
\subset \Omega \simeq U(2)\ .
\eqn\chprogr
$$
The parameter space of the
subgroup $U(1) \times U(1)$ for $\Omega_{ P}$ is
a torus $S^1 \times S^1$ (see Fig.2) which is,
of course,
different from the one corresponding
to the subgroup (\subthree) for $\Omega_R$.
\topinsert
\vskip 1.5cm
\let\picnaturalsize=N
\def\picsize{5cm}
\def\picfilename{f2.epsf}
\input epsf
\ifx\picnaturalsize N\epsfxsize \picsize\fi
\hskip 2.8cm\epsfbox{\picfilename}
\vskip 0.5cm
\abstract{%
{\bf Figure 2.}
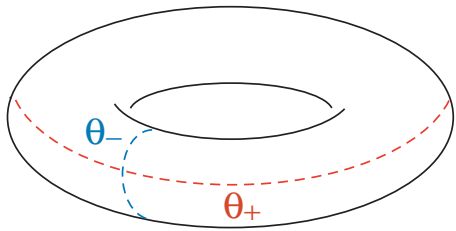
The torus $S^1 \times S^1$ representing the subfamily
$\Omega_{ P}$ in $\Omega$.
}
\endinsert

\medskip
\noindent
{\it 3.1.2. Time-reversal}

We next consider the {\it time-reversal} transformation
defined by
$$
{\cal T}: \quad \varphi(x)  \longrightarrow
({\cal T} \varphi)(x) := \varphi^*(x)\ .
\eqn\timerev
$$
This implies that on vectors we have
$$
\Phi \buildrel {\cal T} \over \longrightarrow \Phi^*\ ,
\qquad
\Phi' \buildrel {\cal T} \over \longrightarrow \Phi'^*\ .
\eqn\vectrev
$$
{}From the boundary condition (\unitrel)
we find that the time-reversal transformation
induces a map on $\Omega$ by
$$
U \buildrel {\cal T} \over \longrightarrow U^{\ssf T}\ ,
\eqn\trmap
$$
where ${\ssf T}$ denotes transposition.
Thus, a given point interaction
is time-reversal invariant
if its characteristic matrix $U$ obeys
$$
U^{\ssf T} = U\ .
\eqn\symumat
$$
This condition is fulfilled if $\beta_R = 0$ in (\stparm), that is,
$$
U = e^{i\xi}
\pmatrix{ \alpha_R + i\alpha_I    & i\beta_{I} \cr
          i\beta_{I}  & \alpha_R - i\alpha_I \cr} \ .
\eqn\tsol
$$
Thus we see that
the time-reversal invariant subspace
$\Omega_{T} \subset \Omega$ is
isomorphic to the coset
space,
$$
\Omega_{T} \simeq {{U(2)}\over{U(1)}}
\simeq S^1 \times S^2\ ,
\eqn\tis
$$
where the $U(1)$ is the subgroup generated by
$\sigma_2$ in the $SU(2) \subset U(2)$.

\medskip
\noindent
{\it 3.1.3. ${\cal PT}$-transformation}

The combined ${\cal PT}$-transformation can
also be considered as
$$
{\cal PT}: \quad \varphi(x)  \longrightarrow
({\cal PT} \varphi)(x) :=
({\cal P}({\cal T} \varphi))(x) = \varphi^*(-x)\ ,
\eqn\ptrev
$$
under which the vectors transform as
$$
\Phi \buildrel {\cal PT}
\over \longrightarrow \sigma_1 \Phi^*\ ,
\qquad
\Phi' \buildrel
{\cal PT} \over \longrightarrow \sigma_1 \Phi'^*\ .
\eqn\veccompt
$$
The map on $\Omega$ induced by the
${\cal PT}$-transformation is then
$$
U \buildrel {\cal PT} \over \longrightarrow
\sigma_1\, U^{\ssf T}\, \sigma_1\ ,
\eqn\ptrmap
$$
The invariant subspace
$\Omega_{PT}$ under the
${\cal PT}$-transformation
is furnished by the set of $U$ obeying the condition,
$$
\sigma_1\, U^{\ssf T}\, \sigma_1 = U\ .
\eqn\cjsymumat
$$
This condition holds for $U$ if
$\alpha_I = 0$ in (\stparm), that is,
$$
U = e^{i\xi} \pmatrix{ \alpha_{R}
                 & \beta_R + i\beta_I \cr
      - \beta_R + i\beta_I & \alpha_{R} \cr} \ ,
\eqn\ptsol
$$
and hence, again, the invariant subspace
$\Omega_{PT} \subset \Omega$ is
isomorphic to the coset,
$$
\Omega_{PT} \simeq {{U(2)}\over{U(1)}}
\simeq S^1 \times S^2\ ,
\eqn\ptis
$$
where now the $U(1)$ is the subgroup
generated by $\sigma_3$.

\medskip
\noindent
{\it 3.1.4. Weyl scaling}

Unlike the previous three
discrete transformations,
the {\it Weyl scaling} transformation
is a continuous transformation with parameter
$\lambda > 0$ and is given by
$$
{\cal W}_\lambda: \quad
\varphi(x) \longrightarrow
({\cal W}_\lambda\varphi)(x) :=
\lambda^{1\over 2} \varphi(\lambda x)\ .
\eqn\weyl
$$
This implies
$$
\varphi(0_\pm) \buildrel {\cal W}_\lambda \over
\longrightarrow \lambda^{1\over 2} \varphi(0_\pm)\ ,
\qquad
\varphi'(0_\pm) \buildrel {\cal W}_\lambda \over
\longrightarrow \lambda^{3\over 2} \varphi'(0_\pm)\ ,
\eqn\wlpty
$$
and hence
$$
\Phi \buildrel {\cal W}_\lambda \over
\longrightarrow \lambda^{1\over 2} \Phi \ ,
\qquad
\Phi' \buildrel {\cal W}_\lambda \over
\longrightarrow \lambda^{3\over 2} \Phi' \ .
\eqn\matpty
$$
The boundary condition (\unitrel)
proves to be unchanged under the Weyl
scaling transformation (\weyl)
if each of the two terms in
(\unitrel) vanishes separately,
$$
(U - I)\Phi = 0\ , \qquad (U + I) \Phi' = 0\ .
\eqn\twobc
$$
Since $\Phi$ and $\Phi'$ cannot vanish simultaneously,
we observe that $U = \pm I$ or else
the two eigenvalues of
$U$ are $+1$ and $-1$, namely,
$$
\det (U - I) = \det (U + I) = 0\ .
\eqn\decouple
$$
This latter is achieved if
$\xi = \pi/2$ and $\alpha_R = 0$ in (\stparm)
where we have
$$
U = i\pmatrix{ i\alpha_{I}    & \beta_R + i\beta_I \cr
         - \beta_R + i\beta_I  & -i\alpha_{I} \cr} \ .
\eqn\wsol
$$
We find that the subset $\Omega_{W}\subset \Omega$
formed by these
matrices $U$ is isomorphic to the sphere,
$$
\Omega_{W} \simeq
{{U(2)}\over{U(1) \times U(1)}} \simeq S^2\ ,
\eqn\sisp
$$
with the second $U(1)$ being the
subgroup generated by $\sigma_3$.
Together with the isolated points
$ \{U = \pm I\}$,
this continuous subset $\Omega_{W}$
provides the scale invariant subfamily
within $\Omega$.
We can see from (\twobc) that, for
scale invariant point interactions,
the scale parameter $L_0$ drops out from the
boundary conditions as expected.

\bigskip
\noindent
{\bf 3.2. Spectrum-preserving $su(2)$}
\medskip

We have seen that the parity transformation
induces a map on $\Omega$ which is
the conjugation of $U$ by $\sigma_1$.
We now provide
transformations which induce
similar maps on $\Omega$ by means of
the conjugation of $U$ by other Pauli
matrices $\sigma_i$ for $i = 2$ and 3.

The first is the {\it half-reflection}
transformation ${\cal R}$ defined by
$$
{\cal R}: \quad \varphi (x)
\longrightarrow
({\cal R}\varphi)(x) := [\Theta(x) -
\Theta(-x)]\varphi(x)\ ,
\eqn\rtrsf
$$
where $\Theta(x)$ is the Heaviside step function.
On boundary vectors, this leads to
$$
\Phi
\buildrel {\cal R} \over \longrightarrow \sigma_3
\Phi\ ,
\qquad
\Phi'
\buildrel {\cal R} \over \longrightarrow \sigma_3
\Phi'\ .
\eqn\vecrtr
$$
We then see in (\unitrel) that
this induces a map on $\Omega$ by
$$
U \buildrel {\cal R} \over \longrightarrow
\sigma_3\, U\, \sigma_3\ .
\eqn\halfrefonu
$$
Hence, the subfamily $\Omega_{R}$ of point
interactions invariant under the half-reflection
transformation is characterized by those
$U$ obeying
$$
\sigma_3\, U \, \sigma_3 = U\ .
\eqn\pinv
$$
The general solution of this is
$$
U = e^{i \xi} \pmatrix{ e^{i \rho} & 0 \cr 0 & e^{- i \rho}
 \cr} \ ,
\eqn\crcmatr
$$
with $\xi \in [0, \pi)$ and
$ \rho \in [0, 2 \pi) $, which is the same as
the one given in (\usep).
This implies
that the subfamily $\Omega_{R}$ is
in fact the one given in (\subthree)
which is the set of interactions missing
in the local description.

The remaining transformation corresponding to
$\sigma_2$ is furnished
by
$$
{\cal Q}: \quad \varphi (x)
\longrightarrow
({\cal Q}\varphi)(x) :=
 i[\Theta(-x) - \Theta(x)]\varphi(-x)\ .
\eqn\qtrsf
$$
This is just the combination
${\cal Q}=-i{\cal R}{\cal P}$, and on vectors
it implements the transformation,
$$
\Phi
\buildrel {\cal Q} \over \longrightarrow \sigma_2
\Phi\ ,
\qquad
\Phi'
\buildrel {\cal Q} \over \longrightarrow \sigma_2
\Phi'\ .
\eqn\qonvec
$$
It follows from (\unitrel) that the induced
map on $\Omega$ is given by
$$
U \buildrel {\cal Q} \over \longrightarrow
\sigma_2\, U\, \sigma_2\ ,
\eqn\qonu
$$
as required.
The subfamily $\Omega_{Q}$ of point
interactions invariant under this
transformation is characterized by
$$
\sigma_2\, U \, \sigma_2 = U\ .
\eqn\qpinv
$$
As is clear now, this admits the following solution,
$$
U = U(\omega_+, \omega_-)
  = e^{i(\omega_+P^+_2 + \omega_-P^-_2)} \ ,
\eqn\qchiralproj
$$
with
$\omega_\pm \in [0, 2\pi)$ and the
projection matrices
$P^\pm_2$ in (\prjthree).
The invariant subfamily
$\Omega_{Q}$ is therefore given by
$$
\Omega_{Q} \simeq U(1) \times U(1)
\subset \Omega \simeq U(2)\ .
\eqn\qchprogr
$$
To sum up, we have altogether three different
tori $S^1 \times S^1$ for $\Omega_{ P}$,
$\Omega_{Q}$
and $\Omega_{R}$ in $\Omega \simeq
S^1 \times S^3$.  These tori are formed by distinct
subgroups $U(1) \times U(1)$ in $U(2)$, where
the first $U(1)$ accounts for the overall
phase factor while
the second $U(1)$ is the subgroup
generated by $\sigma_i$
for $i = 1$, 2, 3.

A remarkable point to note is that the three
discrete transformations, ${\cal P}$, ${\cal Q}$
and ${\cal R}$, satisfy the relations,
$$
{\cal P}{\cal Q}
= - {\cal Q}{\cal R} = i{\cal R}\ , \qquad
{\cal Q}{\cal R}
= - {\cal R}{\cal Q} = i{\cal P}\ , \qquad
{\cal R}{\cal P}
= - {\cal P}{\cal R} = i{\cal Q}\ , \qquad
\eqn\algerel
$$
and
$$
{\cal P}^2 = {\cal Q}^2 = {\cal R}^2 = 1\ .
\eqn\normrel
$$
As a result, the set
$\{{\cal P}, {\cal Q}, {\cal R}\}$ forms an
$su(2)$ algebra.  The vectors $\Phi$ and $\Phi'$
provide distinctive
two-dimensional bases on which the
$su(2)$ is represented in terms of the Pauli
matrices.  In fact, the relations
(\algerel) and (\normrel) are precisely
those fulfilled by the Pauli
matrices.

Importantly,
since the three discrete transformations
$\{{\cal P}, {\cal Q}, {\cal R}\}$
of $su(2)$ act as multiplications by constant factors
along the positive and the negative half lines
and/or the mirror reflection, they are
intrinsically {\it spectrum-preserving}.
More explicitly,
if $\varphi$ is an
energy eigenstate
$
H\, \varphi(x) = E\, \varphi(x)
$
of the Hamiltonian $H$ in (\ham),
then, {\it e.g.},
the state ${\cal R} \varphi$ is also
an eigenstate
of the Hamiltonian with the same energy $E$:
$$
 H\, ({\cal R}\varphi)(x) =
E\, ({\cal R}\varphi)(x)\ .
\eqn\dualegneq
$$
Moreover, if $\varphi$ is a simultaneous
eigenstate of ${\cal P}$ as
${\cal P}\varphi = \pm \varphi$, then
the mapped state has the opposite
parity,
$$
{\cal P}\,({\cal R}\varphi) =
- {\cal R}\,({\cal P}\varphi)
= \mp {\cal R}\varphi\ .
\eqn\opartyrel
$$
Obviously, the same holds for other
generators of $su(2)$ as well.  However,
this does not necessarily
mean that the $su(2)$ is a symmetry,
because the transformations alter the domain $D(H)$ of
the Hamiltonian, according to (\parityonu),
(\halfrefonu) and (\qonu). As we shall see shortly,
there is only a special subclass of domains
characterized by point interactions which remain
unaltered under these transformations. When this
happens, the $su(2)$ becomes a symmetry of the system,
and this $su(2)$ symmetry will be important to
realize the duality discussed in the next section.

We have learned that, in the entire family
$\Omega = U(2)$
of point interactions allowed in one
dimension, there are several subfamilies
characterized by their symmetries.
Before analyzing the structure of the
invariant subfamilies in detail, we
mention here that
these subfamilies can be used to provide
a classification of point interactions.
To see this, let
$U \in U(2)$ be
the characteristic matrix
of a generic point interaction.  Because of the
complementary structure
of the two invariant subspaces
$\Omega_{W}$ and
$\Omega_{R}$, as seen in (\sisp)
and (\subthree), the matrix $U$ can be decomposed
as
$$
U = U_{W}\, U_{R}\ ,
\eqn\decone
$$
where $U_{W}$
is of the form (\wsol) while
$U_{R}$ is given by the form (\crcmatr).
Namely, a generic point interaction is
a \lq product\rq{} of two point interactions,
one is invariant under the
Weyl scaling ${\cal W}_\lambda$
and the other
is invariant under
the half-reflection ${\cal R}$.
More precisely, one can observe that
the decomposition (\decone) provides
a double covering of the entire family
$\Omega \simeq U(2)$.  This can be confirmed
explicitly by computing the product as
$$
U_W\, U_R = e^{i \xi'}
\pmatrix{ \alpha' & \beta' \cr
- {\beta'}^* & {\alpha'}^*
\cr }\ ,
\eqn\uformdec
$$
where
$ \xi' = \xi + \pi / 2 \; (\mod \pi)$,
$ \alpha' = i \alpha_I e^{i\rho} $ and
$ \; \beta' = \beta e^{- i \rho} $.
Conversely, to a given $U$ as the r.h.s.~of
(\uformdec), the parameters
in the decomposition are identified as $ \rho = \arg \alpha'
- \pi / 2 \; (\mod 2
\pi)$, $ \alpha_I = |\alpha'| \geq 0 $ and
$\beta = \beta' e^{i \rho} $.
Thus we need only non-negative
$\alpha_I$ to cover the
$U(2)$ once, and
negative $\alpha_I$ provide another covering of $U(2)$.
{}From this consideration we learn that, once the choice of
the covering is made,
the decomposition $ U = U_W U_R $ is unique
except for the case $\alpha' = 0$ where $\rho$
can be chosen arbitrarily.

Similar decompositions are also
available for other choice of invariant
subfamilies based on the {\it local}
decomposition of $U(2) \simeq S^1 \times S^3$ by
$(S^1 \times S^2) \times S^1$.  For instance,
one may consider the Abelian subgroup
$U(1) \subset \Omega_{R}$
generated by $\sigma_3$, whereby
obtain the decomposition
$U = U_3\, U_{PT}$
for any $U \in U(2)$ using
$U_3$ and $U_{PT}$ from
the $U(1)$
and $\Omega_{PT}$, respectively.
These decompositions
rest basically on the choice of the spheres $S^2$
in $U(2)$, which are obtained, {\it e.g.}, by
the condition analogous to (\decouple),
$$
\det (U - \sigma_i) = \det (U + \sigma_i) = 0\ .
\eqn\gendecouple
$$
Indeed, this leads to the $S^2$ in $\Omega_{PT}$
for $i = 3$ and the one
in $\Omega_{T}$ for $i = 2$.
We also mention that, albeit being local,
the transfer matrix
formalism admits a similar classification, a similar
decomposition of the family $\Omega$, a decomposition
which is physically more sensible in the
sense that the sequence of
matrices in the decomposition from left to right
corresponds to the actual
sequence of point interactions placed
from left to right at the singularity [\CSa].

\bigskip
\noindent
{\bf 3.3. Parity invariant subfamily and spin}
\medskip

We now take a closer look at some of
the invariant subfamilies discussed so far.
We first analyze the parity invariant
subfamily $\Omega_P$ given by (\chiralproj).
For this, recall that any state
$\varphi$ in the Hilbert space (\hsp)
can be decomposed into the sum
$\varphi = \varphi_+ + \varphi_-$
of a symmetric
$\varphi_+$ and antisymmetric states
$\varphi_-$. {}From
$\varphi_+(-x) = \varphi_+(x)$ and
$\varphi_-(-x) = -\varphi_-(x)$ we
obviously have
$$
\varphi_\pm(0_+) = \pm \varphi_\pm(0_-) \ ,
\qquad
\varphi_\pm'(0_+) = \mp \varphi_\pm'(0_-)\ .
\eqn\bdboth
$$
Then, plugging the decomposition into
the boundary condition (\unitrel) and
using the relations (\bdboth), we find
that the condition splits into two conditions,
$$
\eqalign{
\sin{{\theta_+}\over 2}\,
\varphi_+(0_+) + L_0
\cos{{\theta_+}\over 2}\,
\varphi'_+(0_+) &= 0\ ,
\cr
\sin{{\theta_-}\over 2}\,
\varphi_-(0_+) + L_0
\cos{{\theta_-}\over 2}\,
\varphi'_-(0_+) &= 0\ .
}
\eqn\pbcwell
$$
Thus we see that the two angles $\theta_\pm$
specify the connection conditions
at the gap $x = 0_\pm$ separately for symmetric
and antisymmetric states
$\varphi_\pm$.  An important consequence of this
is that the energy spectrum of the symmetric sector is
determined solely by the angle $\theta_+$,
and in the antisymmetric sector by $\theta_-$ alone.
The equivalence of the connection condition
of the two sectors in (\pbcwell) implies
that the spectra are given by the same function
of the angles $\theta_\pm$.

At this point it is worth mentioning
that, among the one-dimensional parity invariant
systems on a line $\R$,
there are two typical singular
potentials which we are familiar with.
One is the delta-function potential,
$V(x) = \delta(x; \theta_+) = c(\theta_+)\,\delta(x)$,
which gives rise
to a gap in the derivative $\varphi'(x)$
of a wave function
$\varphi(x)$ at $x = 0$ proportional to the constant
$c(\theta_+)$ while keeping the continuity of the value
$\varphi(x)$ at $x = 0$.  The other, which is less
familiar, is the epsilon-function potential,
$V(x) = \varepsilon(x; \theta_-)$, which brings about
a gap in the value $\varphi(x)$ at $x = 0$
preserving the continuity in the derivative
$\varphi'(x)$ there.

We may view our system with point interaction
defined on $\R\setminus\{0\}$ as a system
on $\R$ with some effective singular potential.
Conversely, the system on $\R$ with a
delta-function (or epsilon-function) potential
may be regarded as a special case of our systems
on $\R\setminus\{0\}$.
In fact, the boundary conditions implied by
the delta-function potential arises precisely at
$(\theta_+, \theta_-) = (\theta_+, \pi)$.
It is also easy to confirm that the
epsilon-function potential is reproduced at
$(\theta_+, \theta_-) = (0,\theta_-)$.
In particular, at the intersection $(0, \pi)$
we obtain $\varphi_+'(0_+) = \varphi_-(0_+) = 0$,
which implies the smooth continuity both
in the values and the derivatives,
$\varphi(+0) = \varphi(-0)$ and
$\varphi'(+0) = \varphi'(-0)$.  Hence,
the point
$(0, \pi)$ yields nothing but the free system.

\topinsert
\vskip 1.5cm
\let\picnaturalsize=N
\def\picsize{7cm}
\def\picfilename{f3.epsf}
\input epsf
\ifx\picnaturalsize N\epsfxsize \picsize\fi
\hskip 1.5cm\epsfbox{\picfilename}
\vskip 0.5cm
\abstract{%
{\bf Figure 3.}
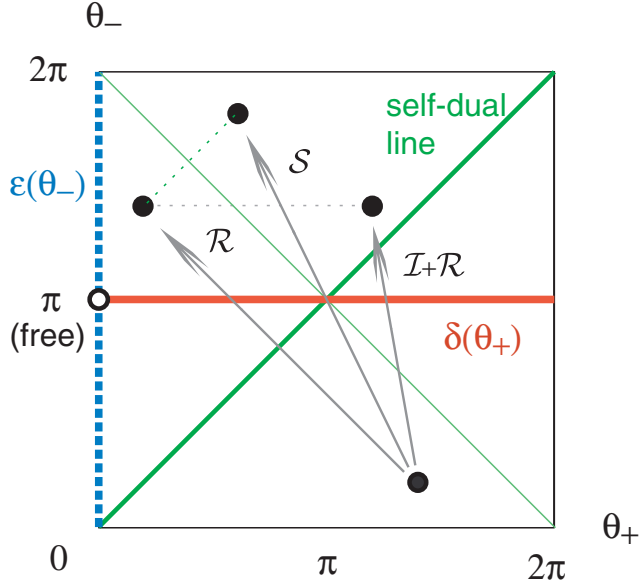
The dissected torus $\Omega_P$.  The ${\cal R}$
transformation and two 
other discrete transformations, ${\cal I}_+{\cal R}$
and ${\cal S}$
mentioned in Sect.4, are
indicated by the arrows.
The horizontal solid line and the vertical dotted line
represent the
delta-function interactions and
the epsilon-function interactions, respectively.
}
\endinsert
Now we ask ourselves what happens if we implement
other transformations of $su(2)$ in the
parity invariant subfamily $\Omega_P$.
Consider the half-reflection ${\cal R}$
applied to the system of parity invariant
point interactions. {}From (\halfrefonu)
we observe that
$$
U(\theta_+, \theta_-)
\buildrel {\cal R} \over \longrightarrow
\sigma_3\, U(\theta_+, \theta_-)\, \sigma_3
= U(\theta_-, \theta_+)\ .
\eqn\dtrsfumatd
$$
The half-reflection
${\cal R}$ therefore induces on $\Omega_P$
the interchange of parameters,
$(\theta_+, \theta_-)
\buildrel {\cal R} \over \longrightarrow
(\theta_-, \theta_+)$, which is
the map across the diagonal
line $\theta+ = \theta_-$ as shown in the
diagram in Fig.3.  On account of
the dual aspect of
the spectrum under the map
as seen in (\dualegneq)
and (\opartyrel) before,
we hereafter call
the two points $(\theta_+, \theta_-)$ and
$(\theta_-, \theta_+)$ the {\it dual} of each
other.

The set of
points in $\Omega_P$ lying on the diagonal line,
{\it i.e.}, the {\it self-dual} points,
forms the subgroup,
$$
\Omega_{SD} \simeq U(1) \subset \Omega_{P}
\simeq U(1) \times U(1)\ ,
\eqn\sdsst
$$
consisting of the matrices,
$$
U
 =e^{i\theta }
  \left( \matrix{
           1&0\cr
           0&1\cr}
  \right)
\ ,
\qquad \theta \in [0, 2\pi)\ .
\eqn\sdsubfamily
$$
{}From the foregoing argument
we see that points on
the diagonal line in $\Omega_{SD}$
are left unchanged under
the half-reflection ${\cal R}$, which
means that the domain $D(H)$ of the
Hamiltonian defined at a point on
$\Omega_{SD}$ is invariant under ${\cal R}$.
Moreover, since ${\cal Q}$ is
just a product of ${\cal P}$
and ${\cal R}$, we conclude that the entire
$su(2)$ arises as a symmetry algebra
on $\Omega_{SD}$,
$$
[H, {\cal P}] = [H, {\cal Q}] = [H, {\cal R}] = 0\ .
\eqn\fulsym
$$
It follows that, for point interactions
defined at self-dual points,
every energy eigenstate
falls into an irreducible
representation of $su(2)$, {\it
i.e.}, it possesses a {\it spin} with respect to
the $su(2)$ algebra of
the generators for
the discrete transformations.
More specifically, because
${\cal R}$ flips the parity of the state,
there always be a double (or even) degeneracy
in every level of energy spectrum on $\Omega_{SD}$,
and this number of degeneracy is given
by the spin of the state.

\bigskip
\noindent
{\bf 3.4. Generic invariant subfamily of $su(2)$}
\medskip

We now argue that
the structure of the invariant
subfamily discussed for $\Omega_P$ is generic,
that is, analogous results can also hold
for other invariant subfamilies $su(2)$.
To see this, let ${\cal X}_i$, $i = 1$, 2, 3, be
the generators $\{ {\cal P}, {\cal Q}, {\cal R}\}$
of the $su(2)$.  Since
${\cal X}_i^2 = 1$ as seen in (\normrel), one can
decompose
the Hilbert space ${\cal H}$ into
the eigenspaces ${\cal H}_\pm$
with eigenvalues $\pm 1$,
$$
{\cal H} = {\cal H}_+ \oplus {\cal H}_-\ ,
\eqn\hlrdec
$$
that is, any state $\varphi \in {\cal H}$ can be
put
$$
\varphi = \varphi_+ + \varphi_-\ ,
\qquad
\varphi_\pm :=
\left({{1 \pm {\cal X}_i}\over 2}\right)\varphi\ .
\eqn\decstates
$$
The corresponding decomposition
of the boundary vectors is
$$
\Phi = \Phi_+ + \Phi_-\ ,
\qquad
\Phi' = \Phi_+' + \Phi_-'\ ,
\eqn\vecdec
$$
where
$$
\Phi_\pm := P^\pm_i \Phi\ ,
\qquad
\Phi_\pm' := P^\pm_i \Phi'\ .
\eqn\prodefgen
$$
Since
the ${\cal X}_i$ transformation
yields
$
\Phi
\buildrel {\cal X}_i \over \longrightarrow
\sigma_i \Phi
$
and
$
\Phi'
\buildrel {\cal X}_i \over \longrightarrow
\sigma_i \Phi'
$,
and since
$$
\sigma_i \Phi_\pm = \pm \Phi_\pm\ ,
\qquad
\sigma_i \Phi_\pm' = \pm \Phi_\pm' \ ,
\eqn\genprhs
$$
we find that $\Phi_\pm$ and $\Phi_\pm'$
are indeed the boundary vectors for the eigenstates
$\varphi_\pm$.  Note that
the matrix form available in
the subfamily invariant
under the ${\cal X}_i$ transformation reads
$$
U = U(\vartheta_+, \vartheta_-)
  = e^{i(\vartheta_+P^+_i + \vartheta_-P^-_i)} \ ,
\eqn\genchiralproj
$$
where $\vartheta_\pm \in [0, 2\pi)$
are two angle parameters.
Then, upon using the identity,
$$
U(\vartheta_+, \vartheta_-) \pm 1
= (e^{i\vartheta_+} \pm 1) P_i^+
+ (e^{i\vartheta_-} \pm 1) P_i^-\ ,
\eqn\combtn
$$
one can
split the boundary condition (\unitrel)
into two conditions imposed on
the eigenstates,
$$
\eqalign{
\sin{{\vartheta_+}\over 2}\,
\Phi_+ + L_0
\cos{{\vartheta_+}\over 2}\,
\Phi'_+ &= 0\ ,
\cr
\sin{{\vartheta_-}\over 2}\,
\Phi_- + L_0
\cos{{\vartheta_-}\over 2}\,
\Phi'_- &= 0\ .
}
\eqn\vecgenbcwell
$$

By construction, each of the projected boundary
vectors has only one independent component.
For example, for ${\cal X}_1 = {\cal P}$
from (\prodefgen) we have
$$
\Phi_+ =
(\varphi_+(0_+), \varphi_+(0_+))^{\ssf T}\ ,
\qquad
\Phi_- =
(\varphi_-(0_+), \varphi_-(0_+))^{\ssf T}\ ,
\eqn\compvecp
$$
(and similarly for the derivatives), and
hence the conditions (\vecgenbcwell)
reduce to (\pbcwell) which we have
obtained earlier.
On the other hand, for ${\cal X}_3 = {\cal R}$
we have
$$
\Phi_+ =
(\varphi_+(0_+), 0)^{\ssf T}\ ,
\qquad
\Phi_- =
(0, \varphi_-(0_-))^{\ssf T}\ ,
\eqn\compvecr
$$
where now the eigenstates $\varphi_\pm$ are
those which have supports only on the left
half line $\R^-$ and the right half line $\R^+$,
respectively.  As a result,
the boundary conditions (\vecgenbcwell) read
$$
\eqalign{
\sin{{\phi_+}\over 2}\,
\varphi_+(0_+) + L_0
\cos{{\phi_+}\over 2}\,
\varphi_+'(0_+) &= 0\ ,
\cr
\sin{{\phi_-}\over 2}\,
\varphi_-(0_-) + L_0
\cos{{\phi_-}\over 2}\,
\varphi_-'(0_-) &= 0\ .
}
\eqn\bcwell
$$
An important point to note
is that these conditions (\bcwell) are nothing
but the ones which
arise when we require the
probability conservations $j(0_+) = j(0_-) = 0$
separately on the two half lines, $\R^+$ and
$\R^-$.  This implies that the subfamily
$\Omega_R$ of point interactions invariant
under the half-reflection ${\cal R}$
describes (non-unique) perfect walls at $x = 0$
through which no probability flow can penetrate,
separating the left $\R^-$
and the right half lines $\R^+$ completely.
For this reason, the subfamily $\Omega_R$
may also be called the {\it separated subfamily}.
Finally, for the subfamily invariant under
${\cal X}_2 = {\cal Q}$ we have
$$
\Phi_+ =
(\varphi_+(0_+), i\varphi_+(0_+))^{\ssf T}\ ,
\qquad
\Phi_- =
(\varphi_-(0_+), i\varphi_-(0_+))^{\ssf T}\ ,
\eqn\compvecq
$$
where $\varphi_\pm$ are symmetric
and antisymmetric (with phase-twisted)
eigenstates of the
${\cal Q}$ transformation.
Thus, in $\Omega_Q$ the boundary conditions
(\vecgenbcwell) become
$$
\eqalign{
\sin{{\omega_+}\over 2}\,
\varphi_+(0_+) + L_0
\cos{{\omega_+}\over 2}\,
\varphi_+'(0_+) &= 0\ ,
\cr
\sin{{\omega_-}\over 2}\,
\varphi_-(0_+) + L_0
\cos{{\omega_-}\over 2}\,
\varphi_-'(0_+) &= 0\ ,
}
\eqn\qbcwell
$$
which are formally identical to
the ones (\pbcwell) in $\Omega_P$.

Returning to the generic invariant
subfamily of the
$su(2)$ transformation ${\cal X}_i$,
we can also see how other spectrum-preserving
transformations
in the $su(2)$ act on the subfamily.
For this we just observe that
the transformation
${\cal X}_j$ with $j \ne i$
induces the map on the characteristic matrix
(\genchiralproj) as
$$
U(\vartheta_+, \vartheta_-)
\buildrel {\cal X}_j \over \longrightarrow
\sigma_j\, U(\vartheta_+, \vartheta_-)\,
\sigma_j
= U(\vartheta_-, \vartheta_+)\ .
\eqn\gendtrsfumatd
$$
The dual map
$(\vartheta_+, \vartheta_-)
\buildrel {\cal X}_j \over \longrightarrow
(\vartheta_-, \vartheta_+)$ therefore arises
generically in the $su(2)$ invariant
subfamily.  Thus, at self-dual points
the $su(2)$ becomes a symmetry of the system.
The set of these self-dual points
is always given by $\Omega_{SD}$ in (\sdsst)
irrespective of the invariant subfamily one
chooses in the $su(2)$.

\ve
\secno=4 \meqno=1


\centerline
{\bf 4. Duality and Anholonomy}
\medskip

Due to the spectrum-preserving $su(2)$ found
in the previous section, the system exhibits
certain
remarkable properties of the energy spectrum.
In this section we discuss two notable aspects
of them, that is, the strong vs. weak duality
and the spectral anholonomy.  The former is
the duality of two distinct systems having
reciprocal coupling constants (one is strong
while the other is weak).  The latter
refers to the global structure in the spectral
flow which resembles those of the geometric (Berry)
phase.  The possibility of interpreting
the system as
a supersymmetric model
will also be mentioned at the end.  For definiteness,
we confine ourselves to
the parity invariant subspace $\Omega_P$.

\bigskip
\noindent
{\bf 4.1. Strong vs. weak duality}
\medskip

Prior to the discussion of the
strong vs. weak coupling duality,
we need to furnish a definition
of the coupling constants for point
interactions in the first place.
In general, a coupling constant is defined
to signify the strength of the interaction
in action, and it is usually chosen so that it
vanishes when the interaction ceases to work.

In the parity invariant subspace $\Omega_P$,
we have seen in (\pbcwell) that the
two angles $\theta_+$ and $\theta_-$ are parameters
which characterize independently the two
sectors, namely, the sector of symmetric
states $\varphi_+$ and the one of
antisymmetric states
$\varphi_-$.  Further, the point
$(\theta_+, \theta_-) = (0, \pi)$
is shown to correspond to
the free system. {}From these, we recognize that
for each of the
two sectors
the coupling constants for parity invariant point
interactions may be defined by two functions
$g_+(\theta_+)$ and $g_-(\theta_-)$
fulfilling
$g_+(0) = 0$ and $g_-(\pi) = 0$.  In what follows,
we shall adopt the simple reciprocal choice
for the two coupling constants,
$$
g_+(\theta_+) :=  \tan {\theta_+\over 2}\ ,
\qquad
g_-(\theta_-) := \cot {\theta_-\over 2}\ .
\eqn\chiralcc
$$
For illustration, we note that under (\chiralcc)
the delta-function
interactions $V(x) = c(\theta_+)\,\delta(x)$ that
appear along the line $(\theta_+, 0)$ have
$c(\theta_+) = -\frac{\hbar^2}{m L_0}
g_+(\theta_+)$, and we shall
regard $g_+(\theta_+)$ as the coupling constant
for the delta-function
interactions.
Analogously, the strength of the
epsilon-function interactions
$V(x) = \varepsilon(x; \theta_-)$
can be specified by $g_-(\theta_-)$.
The further advantage of the definition (\chiralcc)
is that it allows us to
elucidate the duality discussed below in a simple
manner thanks to the identities,
$$
g_+(\theta) = {1\over{g_-(\theta)}}\ ,
\qquad
g_+(\theta \pm \pi) = -{g_-(\theta)}\ .
\eqn\ccprop
$$
However, any other choice for the two functions
would equally serve our purposes, as long as they
are single-valued and monotonous functions of $\theta$
on $\Omega_P$.

We now recall that the half-reflection
${\cal R}$ induces the interchange
of the parameters (\dtrsfumatd)
in $\Omega_P$.  Thus, for
coupling constants given in (\chiralcc),
we observe
$$
(g_+(\theta_+), g_-(\theta_-))
\buildrel {\cal  R} \over
\longrightarrow
(g_+(\theta_-), g_-(\theta_+))
= (1/g_-(\theta_-), 1/g_+(\theta_+))\ .
\eqn\swduality
$$
The transformation ${\cal R}$ is therefore
seen to implement the combination of
exchange and inversion for the
coupling constants in the two sectors.
This implies that the duality of
energy spectrum between
two systems connected by
${\cal R}$ can also be
regarded as the duality between two
systems with strong and weak coupling constants,
namely, it is a {\it strong vs. weak duality}
(see Fig.4).  A typical duality
appears at $\theta_+ = \theta_- \pm \pi$,
for which (\swduality) becomes
$$
(g_+(\theta_+), g_-(\theta_-))
\buildrel {\cal  R} \over
\longrightarrow
(- 1/g_+(\theta_+), -1/g_-(\theta_-))\ .
\eqn\spswduality
$$
\topinsert
\vskip 0.5cm
\let\picnaturalsize=N
\def\picsize{5cm}
\def\picfilename1{f4a.epsf}
\input epsf
\ifx\picnaturalsize N\epsfxsize \picsize\fi
\hskip 4.0cm\epsfbox{\picfilename1}
\vskip 1.5cm
\let\picnaturalsize=N
\def\picsize{7cm}
\def\picfilename2{f4b.epsf}
\input epsf
\ifx\picnaturalsize N\epsfxsize \picsize\fi
\hskip 2.0cm\epsfbox{\picfilename2}
\vskip 0.5cm
\abstract{%
{\bf Figure 4.}
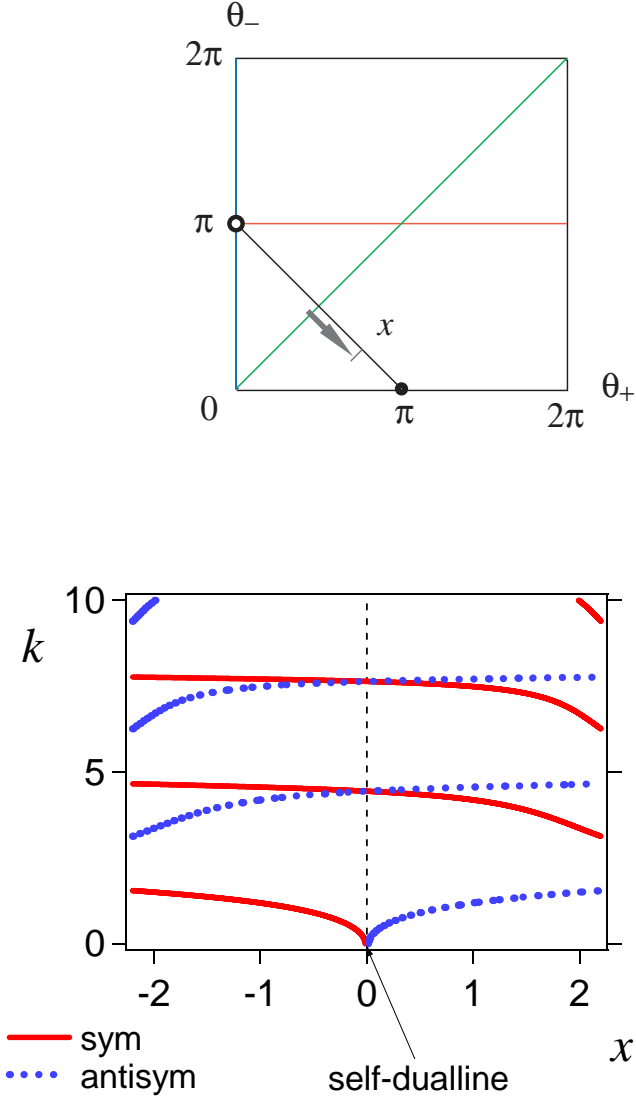
Duality under ${\cal R}$ --- the eigenvalues $k$ are
exactly the same (with the symmetric and antisymmetric
states interchanged) at two points of equal distance
$x$ from the self-dual line.  
The eigenvalues are evaluated for the 
anti-diagonal line connecting the free point
$(0, \pi)$ and $(\pi, 0)$ in the
dissected torus under the Dirichlet boundary conditions,
$\varphi_\pm(l)$
$=\varphi_\pm(-l)$
$=0$ for some $l$ (see Sect.4.2.1). 
}
\endinsert

It is also possible to realize systems with
partial duality
which is the duality between
the symmetric sector in one system and
the antisymmetric sector in the other, but not
necessarily reversely.
For this we need
the translations ${\cal I}_\pm$ by a half-cycle
in the two angles $\theta_\pm$,
$$
\eqalign{
U(\theta_+, \theta_-)
\buildrel {\cal  I}_+ \over
\longrightarrow e^{i\pi P_1^+}\,
U(\theta_+, \theta_-)
&= U(\theta_+ \pm \pi, \theta_-) \ , \cr
U(\theta_+, \theta_-)
\buildrel {\cal  I}_- \over
\longrightarrow
U(\theta_+, \theta_-) \, e^{i\pi P_1^-}
&= U(\theta_+, \theta_- \pm \pi) \ ,
}
\eqn\spinv
$$
where the sign of the shift
$\pm \pi$ depends on where
$\theta_\pm$ lie in the allowed
region $[0, 2\pi)$.
Under these,
the coupling constants
become
$$
\eqalign{
(g_+(\theta_+), g_-(\theta_-))
&\buildrel {\cal  I}_+ \over
\longrightarrow
(- 1/g_+(\theta_+), g_-(\theta_-))\ ,\cr
(g_+(\theta_+), g_-(\theta_-))
&\buildrel {\cal  I}_- \over
\longrightarrow
(g_+(\theta_+), - 1/g_-(\theta_-))\ ,
}
\eqn\parinvminus
$$
and hence ${\cal I}_\pm$ implement {\it coupling
inversions} followed by the sign change in each
of the sectors.

By combining ${\cal R}$
and ${\cal I}_\pm$ as
${\cal I}_\pm {\cal R}$
($= {\cal R} {\cal I}_\mp$)
we obtain
$$
\eqalign{
(g_+(\theta_+), g_-(\theta_-))
&\buildrel {\cal  I_+ R} \over
\longrightarrow
(- g_-(\theta_-), 1/g_+(\theta_+))\ ,\cr
(g_+(\theta_+), g_-(\theta_-))
&\buildrel {\cal  I_- R} \over
\longrightarrow
(1/g_-(\theta_-), - g_+(\theta_+))\ .
}
\eqn\partdual
$$
Equivalently, in the angle coordinates on $\Omega_P$
we have
$(\theta_+, \theta_-) \buildrel {\cal  I_+ R} \over
\longrightarrow (\theta_- \pm \pi, \theta_+)$ and
$(\theta_+, \theta_-) \buildrel {\cal  I_- R} \over
\longrightarrow (\theta_-, \theta_+ \pm \pi)$,
under which the energy spectrum in one sector is
preserved in the other sector, showing that
${\cal  I_\pm R}$ are partial duality transformations.
In particular,
we find that ${\cal  I_+ R}$ maps
$(\theta_+, \pi)$ to $(0, \theta_+)$
for $\theta_+ \in [0, 2\pi)$, which
corresponds to sending the
delta-function
interaction to the epsilon-function
interaction,
$$
\delta(x; \theta_+) \buildrel {\cal  I_+ R} \over
\longrightarrow
\varepsilon(x; \theta_+)\ .
\eqn\amex
$$
This is an example of strong vs. weak (semi) duality
because
the coupling constant is inverted
$g_+(\theta_+) \buildrel {\cal  I_+ R} \over
\longrightarrow
1/{g_+(\theta_+)}$
as seen in (\partdual).
This semi-duality has been noticed earlier
in Ref.[\CSb]. 

There are two other discrete transformations
worth mentioning.  One of them is given by
$$
U(\theta_+, \theta_-)
\buildrel {\cal S} \over
\longrightarrow
U^{-1}(\theta_+, \theta_-)
= U(-\theta_+, -\theta_-)\ ,
\eqn\no
$$
which causes the {\it signature change}
in the coupling constants,
$$
(g_+(\theta_+), g_-(\theta_-))
\buildrel {\cal S} \over
\longrightarrow
(-g_+(\theta_+), -g_-(\theta_-))\ .
\eqn\no
$$
The other is the combination
$
{\cal C}
:=
{\cal S}{\cal I}_+{\cal I}_-{\cal R}
=
{\cal R}{\cal I}_+{\cal I}_-{\cal S}
$
which provides the {\it coupling exchange}
of even and odd strengths,
$$
(g_+(\theta_+), g_-(\theta_-))
\buildrel {\cal C} \over
\longrightarrow
(g_-(\theta_-), g_+(\theta_+))\ .
\eqn\no
$$
Unlike the half-reflection
transformation ${\cal R}$,
this does not preserve the spectra.
We mention that
the coupling exchange ${\cal C}$
is in fact induced by the interchange
of $\Phi$ and $\Phi'$ in
the boundary condition (\unitrel),
and that it is related to
the charge conjugation when the system
is reformulated relativistically.

We remark at this point that, in view of the
fact that all information about a point
interaction is encoded in the boundary
condition (\unitrel),
any physical
quantity ${\cal O}$, such as the transmission
rate or the energy levels
obtained under
the point interactions, is a function of
the parameters $\theta_+$ and $\theta_-$
only through the
coupling constants,
${\cal O}
= {\cal O}(g_+(\theta_+), g_-(\theta_-))$.
Thus, dualities similar to the one found
here can always arise if ${\cal O}$ admits
a simple relation with the $su(2)$ elements,
like the one enjoyed by the
Hamiltonian.

\bigskip
\noindent
{\bf 4.2. Anholonomy}
\medskip

\noindent
{\it 4.2.1. Geometric (Berry) phase on $\Omega_W$}

It is well-known that spectral anholonomy
can appear in the presence of spectral
singularity, {\it i.e.}, degenerate points
in the spectral parameter space.
Thus the scale invariant
sphere $\Omega_W \simeq S^2$
which enjoys the isospectral property
may furnish a suitable place to
observe the spectral anholonomy
if there exists some spectral singularity inside the
sphere $S^2$.
Indeed, if one puts the radius of
the sphere zero by setting
$\alpha_I = \beta_R = \beta_I = 0$ while keeping
$\alpha_R = 1$ and $\xi = \pi/2$, one finds that
the corresponding point in $\Omega$ belongs
to $\Omega_{SD}$ where there occurs a double degeneracy
at each level.  The type of spectral anholonomy
we find here
is the geometric
(Berry) phase that arises when one completes
a cycle along a loop
on the sphere.

To discuss it, for definiteness we put
the particle in a box
given, say,
by the interval $-l \le x \le l$,
and place the
Dirichlet condition
$\varphi_\pm(l) =\varphi_\pm(-l) = 0$.
The energy eigenfunction is then given by (see Appendix
B)
$$
    \varphi_k (x) = A_k e^{ikx} + B_k e^{-ikx}\ ,
\eqn\no
$$
with appropriate coefficients $A_k$, $B_k$ and
the momentum $k$ satisfying (\unitrel) which reads
$$
kL_0 \cot{kl} = { { \sin\xi + \sqrt{1 - \alpha_R^2} }
                \over { \cos\xi + \alpha_R } }\ .
\eqn\dspco
$$
On the the scale invariant
sphere $\Omega_W \simeq S^2$ given by
$\alpha_I^2+\beta_R^2+\beta_I^2  = 1$
this is simplified into $\cos{k l} = 0$ and hence
the allowed momenta are
$$
k(\alpha_I, \beta_R, \beta_I)
= \left(n-{1\over 2}\right)\pi\ ,
\qquad
 n=1,2,\ldots
\eqn\no
$$
which are independent of the parameters.
Using the polar coordinates,
$$
\alpha_I = \cos\theta\ ,
\qquad
\beta_R = \sin\theta \cos\phi\ ,
\qquad
\beta_I = \sin\theta \sin\phi\ ,
\eqn\no
$$
the eigenfunction reads
$$
\varphi_k(x;\theta,\phi) = \cos{\theta\over 2}\, \xi_+(x)
+\sin{\theta\over 2} e^{i(\phi+{\pi\over 2})}\, \xi_-(x)\ ,
\eqn\no
$$
where
$$
\xi_\pm(x) := \sqrt{1\over l} \sin k(x\mp l) \Theta(\pm x)\ .
\eqn\no
$$

Let us now take an arbitrary loop ${\cal C}$ on the sphere
and evaluate the phase $e^{i\gamma({\cal C})}$
that the eigenstate acquires when it
completes a cycle along the loop.
According to Berry (see, {\it e.g.}, [\SW] and
references therein)
it is given by
$$
\gamma({\cal C}) = \oint_{\cal C} A\ ,
\eqn\no
$$
where $A$ is the Berry connection 1-form,
$$
A := \left<\varphi \right|
i{\partial\over{\partial \theta}}
\left. \varphi \right>\, d\theta
+ \left<\varphi \right|
i{\partial\over{\partial \phi}}
\left. \varphi \right>\, d\phi
=  -\sin^2{\theta\over 2}\, d\phi\ .
\eqn\no
$$
The curvature 2-form $F$ is then found to be
$$
F = dA = -{1\over 2}\sin\theta\,d\theta d\phi\ ,
\eqn\no
$$
which is precisely the potential for the Dirac
monopole with strength $g = -1$.  The geometric
phase factor $\gamma({\cal C})$ is therefore
the magnetic flux which is given by the magnetic field
$
B_r = - 1/2
$
times the solid angle subtended by the loop.
(A Berry phase similar to this has been mentioned
in [\EG].)

\noindent
{\it 4.2.2. Anholonomy on $\Omega_P$}
\topinsert
\vskip 0.5cm
\let\picnaturalsize=N
\def\picsize{5cm}
\def\picfilename1{f5a.epsf}
\input epsf
\ifx\picnaturalsize N\epsfxsize \picsize\fi
\hskip 4.0cm\epsfbox{\picfilename1}
\vskip 1.5cm
\let\picnaturalsize=N
\def\picsize{7cm}
\def\picfilename2{f5b.epsf}
\input epsf
\ifx\picnaturalsize N\epsfxsize \picsize\fi
\hskip 2.0cm\epsfbox{\picfilename2}
\vskip 0.5cm
\abstract{%
{\bf Figure 5.}
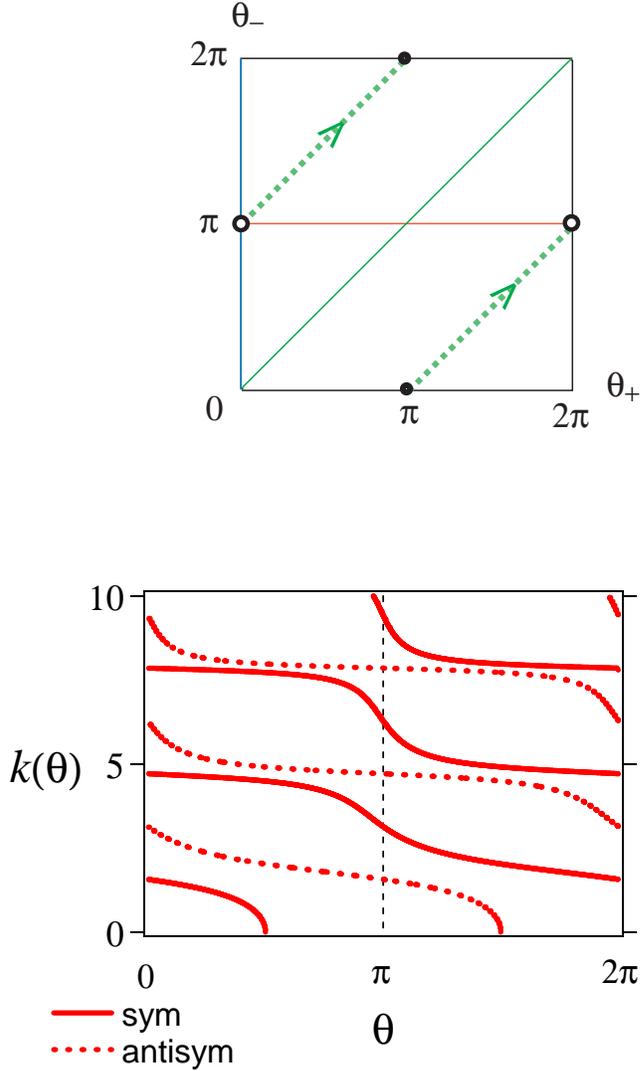
Spiral anholonomy --- each level of $k$ gets
shifted after the cycle 
$\theta$ ($= \theta_+$) 
$\rightarrow$
$\theta + 2\pi$ along the diagonal line 
passing through the
free point $(0, \pi)$ is completed, even though
the spectrum is left unchanged as a whole. 
}
\endinsert

Another type of spectral anholonomy can be
observed in the parity invariant
torus $\Omega_P \simeq S^1 \times S^1$ where
the degenerate, self-dual subfamily $\Omega_{SD}$
is given by a circle that winds the torus
diagonally in the two independent cycles.
Again, we consider the periodic
Dirichlet boundary condition (See Appendix B)
and obtain the spectrum
determined by
$$
k L_0\cot kl = \tan {\theta_+\over 2} \ ,
\qquad
k L_0\cot kl = \tan {\theta_-\over 2} \ ,
\eqn\no
$$
in the symmetric and the antisymmetric
sector, respectively.  Since each of the
momenta is a monotonously decreasing
function of the angle
parameter as
$k = k(\theta_+)$ and $k = k(\theta_-)$,
it is evident that each energy level acquires
a spiral anholonomy as one completes a cycle along
any of the two associated cycles of the torus
$\Omega_P \simeq S^1 \times S^1$.  The double spiral
arises when the cycle is completed simultaneously
in the two parameters $\theta_+$ and $\theta_-$ crossing
the self-dual circle $\Omega_{SD}$,
on account of the fact that even and
odd eigenstates arise alternately in the spectrum
(see Fig.5).
This is
the origin of
the double spiral anholonomy pointed out earlier
in [\CHa].

\bigskip
\noindent
{\bf 4.3. Supersymmetry}
\medskip

As observed in Sect.3.3,
point interactions at
self-dual points in $\Omega_{SD}$
of the form (\sdsubfamily)
give rise to a double (or even) degeneracy at
each level in the spectrum.
Since these degenerate
levels are paired into two states
with opposite parity ${\cal P}$, or
more generally with $\pm$
eigenstates of any of
the $su(2)$ generators, we may ask if the systems
with those point interactions exhibit
some kind of supersymmetry (SUSY).  This issue will
be examined here.

For this we recall briefly the basics of
SUSY quantum mechanics (see, {\it e.g.}, [\Junker]).
A SUSY quantum system is defined by the set
$\{\widehat H,
\widehat Q_1, \ldots, \widehat Q_N; {\cal H}\}$
where $\widehat H$ is the
Hamiltonian operator, $\widehat Q_i$ for
$i = 1, \ldots, N$
are self-adjoint operators called supercharges and
${\cal H}$ is the Hilbert space.  The operators
are subject to the relations,
$$
\bigl\{\widehat Q_i, \, \widehat Q_j\bigr\}
= \widehat H\, \delta_{ij}\ .
\eqn\susyalg
$$
Note that from (\susyalg) it follows that
$\widehat H = 2\widehat Q_i^2$ and
$[\widehat H, \, \widehat Q_i] = 0$ for any $i$.

The scheme of SUSY quantum mechanics which is
most familiar for us is the $N = 2$ Witten model [\Wit].
In the standard formulation on a line $\R$, the model
presupposes the Hilbert space,
$$
{\cal H} = L^2(\R) \otimes\C^2\ ,
\eqn\stdhbt
$$
which is graded by $\C^2$.   The supercharges
are then provided by
$$
\eqalign{
\widehat Q_1 &=
{1\over{\sqrt 2}}\left(-{{i\hbar}\over{\sqrt{2m}}}
{d\over{dx}} \otimes  \sigma_1 + \Lambda(x)
\otimes  \sigma_2 \right)\ ,
\cr
\widehat Q_2 &=
{1\over{\sqrt 2}}\left(-{{i\hbar}\over{\sqrt{2m}}}
{d\over{dx}}\otimes  \sigma_2 - \Lambda(x)
\otimes  \sigma_1 \right) \ ,
}
\eqn\wconst
$$
with a real-valued, continuously
differentiable function $\Lambda(x)$.
The Hamiltonian is then found to be
$$
\widehat H = \left(-{{\hbar^2}\over{2m}}
{{d^2}\over{dx^2}} + \Lambda^2(x) \right)\otimes  1
+ \Lambda'(x)\otimes  \sigma_3 \ .
\eqn\hamsusy
$$
Note that our basis
in the graded space is chosen so that
$\widehat H$ be diagonal
with respect to $\sigma_3$ which is
the element associated with
the half-reflection ${\cal R}$ in the $su(2)$,
but any other choice for the basis works as well.

In order to fit our system with the Witten model,
we need to introduce the graded structure $\C^2$
to our Hilbert space.
In fact, such a structure has already been
equipped with our system under the use of
the two-dimensional
boundary vectors in (\unitrel).
Explicitly, we consider our Hilbert space
(\hsp) to be the sum (\hlrdec) of two eigenspaces
${\cal H}_\pm$ of ${\cal R}$, which are
just ${\cal H}_\pm = L^2(\R^\pm)$.
Since $L^2(\R^+) \simeq L^2(\R^-)$, we have
$$
{\cal H} = L^2(\R^+) \otimes\C^2\ ,
\eqn\grhilbert
$$
where the state decomposition
$\varphi = \varphi_+ + \varphi_-$ for
$\varphi_\pm \in L^2(\R^\pm)$ in
(\decstates)
can now be implemented by means of
the projection operators $P^\pm_3$
for the two-dimensional vector state,
$$
\Phi(x) := (\varphi_+(x), \varphi_-(x))^{\ssf T}\ .
\eqn\decomppsi
$$
At the boundary, the vector state $\Phi(x)$
reduces to the boundary vector $\Phi$ as required,
and this is consistent with the decomposition
(\vecdec) and (\compvecr) discussed earlier.
Accordingly, the projected Hamiltonian $H_\pm$ acting
on the states $\varphi_\pm$ can be obtained
from (\hamsusy) as
$\widehat H = H_+ P_3^+ +  H_- P_3^-$ with
$$
H_\pm = -{{\hbar^2}\over{2m}}
{{d^2}\over{dx^2}} + V_\pm(x)\ ,
\eqn\decham
$$
where
$$
V_\pm(x) =
\Lambda^2(x) \pm
{{\hbar}\over{\sqrt{2m}}}\Lambda'(x) \ .
\eqn\chpot
$$

Thus our question becomes if there exists
a potential $\Lambda(x)$
defined on the
half-line $\R^+$ that can reproduce our Hamiltonian
$H$ in (\ham).  Apparently, this is trivially answered
since if we wish to have $H_\pm = H$ formally
on each of the half-lines $\R^\pm$,
we need $\Lambda(x) = 0$ identically.
However, there still remains a nontrivial aspect
because in our case the characteristics of the
interaction is encoded in the boundary condition
rather than in the formal differential operator.
The real question, therefore,
is whether the boundary
condition (\unitrel) for $U \in \Omega_{SD}$
admits supercharges $\widehat Q_1$ and $\widehat Q_2$
in (\wconst) which are self-adjoint.

This can be answered by observing that,
for $\Lambda(x) = 0$, the supercharges
are basically the momentum operator
$p := -i\hbar{d\over{dx}}$.   The self-adjointness
of the supercharges can thus be examined by
the self-adjointness of $p$.
Evidently, $p$ becomes
self-adjoint in both of ${\cal H}_\pm$
if and only if the wave functions
vanish at the point of interaction,
$\varphi_+(0_+) = \varphi_-(0_-) = 0$.
Since for
$U \in \Omega_{SD}$ the boundary condition
is given by (\bcwell) with
$\phi_+ = \phi_- = \theta \in [0, 2\pi)$,
we see that this is achieved at
$(\theta_+, \theta_-) = (\pi, \pi)$.
{}From our previous argument,
we notice that this corresponds to
the delta-function interaction with
infinite strength $g_+(\theta_+) \to \pm \infty$.

To sum up, we have learned that, among
the self-dual points $(\theta,
\theta)$ for $\theta \in [0, 2\pi)$ where
the $su(2)$ symmetry arises, there exists
an exceptional point
$(\pi, \pi)$ which enjoys
the standard SUSY as a Witten model.
For other points $\theta \ne \pi$, however, the question of
being a SUSY model --- possibly under generalized
SUSY charges other than (\wconst) ---
remains to be explored,
although the double degeneracy indicates that
that is the case in general.

\ve
\secno=5 \meqno=1


\centerline
{\bf 5. Summary and Discussions}
\medskip

In this paper, we have investigated the physical
properties of the quantum mechanical point interaction
in one dimension.
The allowed types of interactions
in one dimension are classified by the parameter
group $U(2)$, in which we
focused on a number of subfamilies characterized
by symmetries.  In particular,
we have presented a detailed analysis of the parity
invariant subfamily $U(1) \times U(1)$
which is the set of left-right symmetric
point interactions.
The coupling constants of the point interaction,
which describe separately
the strengths of the interaction for symmetric and
antisymmetric states,
have been defined by means of the parameters
of the subfamily.  It has been found that
the strong vs. weak duality
observed in the parity invariant subfamily is a
direct consequence of the existence of
spectrum-preserving discrete maps in the subfamily.
More generally, we have shown that
the generators of these discrete
maps form an $su(2)$ algebra, and that any of the
$su(2)$ generators defines an invariant subfamily
in which a similar duality can be observed.
We have mentioned
a distinguished $U(1)$ subfamily, consisting
of interactions invariant under all of
the $su(2)$ transformations, which furnishes
a singular circle
in the parameter space $U(2)$
where
every energy level of states
is doubly degenerate.  Because of this singularity
in the spectral space, one can expect some kind of
quantum anholonomy to arise when one completes
a cycle in the space.  Indeed,
for the parity invariant subfamily
we have found a double spiral structure
of the energy levels,
whereas for the
Weyl scaling invariant subfamily we have observed
an induced magnetic monopole leading to
a geometric (Berry) phase.  Further,
we have pointed out that at one point
in the $U(1)$ subfamily the system
can be regarded as a
Witten model, where the
double degeneracy can be accounted for
in terms of the supersymmetry.

These features,
the strong vs. weak duality, quantum anholonomy and
supersymmetry, are usually associated to
more involved systems of quantum field
theories or string theories.  However, what
we have found in this paper is that they may arise
as generic, not accidental, features of
a vastly simpler setting of one dimensional quantum
mechanical system with a single point defect.
In other words,
the low-energy
single-particle quantum mechanics already possesses
nontrivial characteristics which have not been
widely recognized so far.  This implies that,
if we can
fabricate one dimensional devices with a point defect
whose type of interactions is under our control, it is
possible to observe these exotic physical phenomena at the
laboratory level without invoking the
large facility for high energy experiments.
This is not entirely inconceivable under
the present day progress in nano-scale technology, and
for applications we
may ponder on such things as the quantum interferometer
using quantum wires with point defects, or the
\lq quantum pump' which is the device
that can pump the energy through
a cycle in the parameter space exploiting the
anholonomy of the spectral structure.
In fact,
attempts are currently being made at constructing
the quantum filters mentioned in Appendix B, which
seems to have immediate relevance to the field of
quantum information processing or quantum computing
[\BBBJR].

We end this paper by stating our hope that
our analysis presented here
serve as a basis for
realizing the potential use as well as the physics
of the one
dimensional system with point interaction, which is
seemingly innocent but actually full of intriguing
features especially when it is controllable,
and that effort
be made for making the potential a reality in the
near future.

\bigskip
\noindent
{\bf Acknowledgement:}
T.C.~thanks
Prof.~T.~Shigehara and 
Prof.~T.~D.~Cohen
for helpful discussions.
This work has been supported in part by
the Grant-in-Aid for Scientific Research (C)
(No.~10640301 and No.~11640396) by
the Japanese
Ministry of
Education, Science, Sports and Culture.

\ve

\secno=0 \appno=1 \meqno=1

\centerline{\bf Appendix A}
\bigskip

In Sect.2, we presented an elementary argument which shows
that the point interaction in one dimension is characterized
by the group $U(2)$.  The essence of our argument
was that
the probability conservation (\pccond) --- which is
equivalent to the self-adjointness of the Hamiltonian
(\ham) --- requires the condition (\samemodulo),
that is, the
norm of the two vectors,
$$
\Phi^{(\pm)} := \Phi \pm iL_0 \, \Phi'\ ,
\eqn\colbv
$$
constructed from the boundary vectors in (\vectors) with
some fixed $L_0 \ne 0$ be equal,
$$
\vert \Phi^{(+)} \vert = \vert \Phi^{(-)}\vert\ ,
\eqn\cnor
$$
and, therefore,
these vectors must be related by an unitary matrix
$U \in U(2)$.
Upon a rigorous ground, however,
there are two points which have
remained to be shown.  The first point is that
the matrix $U$ is actually independent of the choice of the
state $\varphi$ used for the vectors in (\vectors).
The second is that the parameter
$L_0$ adds no extra freedom to the variety of
point interactions other than the $U(2)$ group.
For completeness, in this Appendix we wish to
prove these two points, and
thereby fill the remaining gap between our
simple derivation and the
technically more involved one based on the theory of
self-adjoint extensions.

\medskip
\noindent{\bf A.1. State-independence of the matrix $U$}

Before we start, let us observe that, if a state
$\varphi$ which belongs to a self-adjoint domain $D(H)$
of the Hamiltonian operator $H$ has
$\Phi\++ =  \Phi\-- = 0$,
the condition (\cnor)
is trivially fulfilled and no question of
the independence arises.
Thus, in what follows we consider only states for which
$\Phi\++$ and/or
$\Phi\--$ is nonvanishing.  Note that such states must
exist in $D(H)$ because otherwise the domain $D(H)$ would
not be closed as required by the self-adjointness of $H$.

First, we show that there exists a pair of two states
$\varphi_1, \varphi_2 \in
D(H)$ for which the corresponding $\Phi_1\++$ and
$\Phi_2\++$ are
linearly independent, and/or
$\Phi_1\--$ and
$\Phi_2\--$ are linearly independent.  Here
the linear independence is understood by viewing the vectors
in the two dimensional vector space
$\C^2$ which is equipped with the inner product
$ \<\Phi_1 , \Phi_2 \> := \Phi_1^\dagger \Phi_2 $.
Indeed, if such a pair does not exist,
then  all $\varphi \in D(H)$ are such that the corresponding
two
$\Phi\++$s are each other's multiple and, similarly, the
$\Phi\--$s are each other's multiple.  Then it is easy
to find
a smooth function
$\varphi\n$ with
$
| \Phi\n\++  | = | \Phi\n\-- | > 0
$ such that $\Phi\n\++$ is orthogonal to these
$\Phi\++$s and $\Phi\n\--$ is orthogonal to these $\Phi\--$s.
It follows that, for
any $\varphi \in D(H)$ and $ \alpha, \beta \in \C $,
the linear combination
$
\alpha \varphi + \beta \varphi\n
$
will satisfy
    $$
    | \alpha \Phi\++ + \beta \Phi\n\++ | =
    | \alpha \Phi\-- + \beta \Phi\n\-- | \ ,
    \eqn\no
    $$
because
    $$
    \eqalign{
    | \alpha \Phi\++ +  \beta \Phi\n\++ |^2 &=
    |\alpha|^2 | \Phi\++ |^2 + |\beta|^2 | \Phi\n\++ |^2 \cr
    &=
    |\alpha|^2 | \Phi\-- |^2 + |\beta|^2 | \Phi\n\-- |^2 =
    | \alpha \Phi\-- + \beta \Phi\n\-- |^2\ .
    }
    \eqn\lcnormcomp
    $$
This implies that the domain $D(H)$ can be enlarged by
the linear combinations $\alpha \varphi + \beta \varphi\n $
to the bigger domain $ D(H)\n = \overline{ D(H) \oplus
\{ \beta \varphi\n \, | \, \beta \in \C \} } $ on which
the Hamiltonian operator is still
self-adjoint.  However, this is impossible since the domain
of a self-adjoint operator cannot be extended any further
(maximal symmetric), see [\RS].

Having shown the existence of a pair
$\varphi_1, \varphi_2 \in D(H) $ for which
$\Phi_1\++$ and $\Phi_2\++$, and/or
$\Phi_1\--$ and
$\Phi_2\--$, are linearly independent,
we can now find
(via a Gram-Schmidt orthogonalization) a
new pair
$\varphi_1$ and $\varphi_2$ such that
$ \Phi_1\++ \perp \Phi_2\++ $ and/or
$\Phi_1\-- \perp \Phi_2\-- $ holds.
Without loss
of generality we can assume that, {\it e.g.,}
$ \Phi_1\++ \perp
\Phi_2\++$.
But then we can
show that $ \Phi_1\-- \perp \Phi_2\--
$ holds, too.
For this, recall that for $\varphi_1$, $\varphi_2
\in D(H)$ we have
$ \varphi_1 + e^{i \omega} \varphi_2 \in D(H)$
for an arbitrary
$ \omega \in [ 0 , 2 \pi ) $, and hence we have
$ | \Phi_1\++ + e^{i \omega} \Phi_2\++ | =
| \Phi_1\-- + e^{i\omega} \Phi_2\-- | $.
Squaring both sides and using
$ | \Phi_i^{(\pm)} |^2 = | \Phi_i^{(\mp)} |^2 $ for
$i = 1$, 2, we find $ \Re [ e^{i \omega} \< \Phi_1\-- ,
\Phi_2\-- \> ] = 0 $ for any $\omega$ and, consequently,
$ \<\Phi_1\-- , \Phi_2\-- \> = 0 $.

Thus we have established that
there is a pair
$
\varphi_1,\,
\varphi_2
\in D(H)
$ such that
$ \Phi_1\++
\perp \Phi_2\++ $ and $ \Phi_1\-- \perp \Phi_2\-- $.
These vectors can be normalized as
$ | \Phi_1^{(\pm)} | = |\Phi_2^{(\pm)} |  = 1 $
by rescaling the states of the pair appropriately.
{}From this we see that there
exists a unique unitary matrix $ U : \C^2 \to
\C^2 $ such that
    $$
    U \Phi_1\++ = \Phi_1\--, \qquad
    U \Phi_2\++ = \Phi_2\--.
    \eqn\uudef
    $$
It remains to show that
this $U$ is actually universal for $D(H)$, that is,
$$
U \Phi\++ = \Phi\-- \ ,
\qquad
\forall \varphi\in D(H) \ .
\eqn\aimeq
$$

To this end, we note that each of the sets
$\{\Phi_1\++, \Phi_2\++\}$ and $\{\Phi_1\--, \Phi_2\--\}$
forms an orthonormal basis in $\C^2$, and hence
the vectors $\Phi^{(\pm)}$ corresponding to $\varphi$
can be expanded as
    $$
    \Phi\++ = \alpha  \Phi_1\++ + \beta  \Phi_2\++, \qquad
    \Phi\-- = \alpha' \Phi_1\-- + \beta' \Phi_2\--,
    \eqn\expansions
    $$
with some coefficients
$\alpha$, $\beta$, $\alpha'$ and $\beta' \in \C$.
Since the condition (\cnor) must be fulfilled for
any vectors given by a linear
combination of $\varphi_1$,
$\varphi_2$ and $\varphi$, we consider, in particular, the combination
$e^{i \omega_1} \varphi_1 + e^{i \omega_2} \varphi_2 + \varphi$
with $ \omega_1, \omega_2 \in [ 0 , 2 \pi ) $.  For this state,
the condition (\cnor) then reads
    $$
    | e^{i \omega_1} \Phi_1\++ + e^{i \omega_2} \Phi_2\++
    + \Phi\++ |^2 =
    | e^{i \omega_1} \Phi_1\-- + e^{i \omega_2} \Phi_2\--
    + \Phi\-- |^2\ .
    \eqn\no
    $$
Making use
of the expansions (\expansions) and the orthonormality
of the bases, we obtain
    $$
    2 \Re [ e^{-i \omega_1} ( \alpha - \alpha') ] +
    2 \Re [ e^{-i \omega_2} ( \beta  - \beta' ) ] =
    |\alpha'|^2 - |\alpha|^2 + |\beta'|^2 - |\beta|^2.
    \eqn\rere
    $$
Since the right hand side is independent of
the arbitrary parameters $\omega_1$ and $\omega_2$,
this equality can hold if and only if
$ \alpha = \alpha'
$ and $ \beta = \beta' $.  Plugging these into (\expansions)
one obtains the identity (\aimeq) as claimed.

\medskip
\noindent{\bf A.2. $L_0$ adds no extra freedom}

Next we show that the parameter
$L_0$ appearing in (\unitrel)
does not give any additional freedom in characterizing
the boundary conditions other than those given by the
$U(2)$ group.
For this, we first note that any $U(2)$ matrix $U$
can be diagonalized using some appropriate
unitary matrix $V$
as
$$
U \rightarrow VUV^{-1} =
D := \pmatrix{e^{i\mu_+} & 0 \cr 0 & e^{i\mu_-} \cr},
\qquad \mu_+, \, \mu_- \in [0, 2\pi).
\eqn\diagmat
$$
These parameters, $\mu_+$ and $\mu_-$ in $D$, may be regarded
as two of the four parameters of $U$ that arise under the
decomposition $U = V^{-1}DV$.
In fact, as we will see in Appendix B, 
they are related to $\xi$ and $\rho =
\arccos \alpha_R$ as 
$ \mu_\pm = \xi \pm \rho \, \, (\hbox{mod } 2 \pi) $
(cf.\ (B.16)). 
We now define new basis vectors,
$$
\Psi = \pmatrix{\psi(0_+) \cr \psi(0_-) \cr}
:= V\Phi\ ,
\qquad
\Psi' = \pmatrix{\psi'(0_+) \cr - \psi'(0_-) \cr}
:= V\Phi'\ .
\eqn\newbcvec
$$
We note that the components in the new basis
(\newbcvec) are a mixture of the components
of the original basis (\vectors) and hence their arguments
are only symbolic, but they
are still independent of $L_0$.
In terms of the new basis, the boundary
conditions (\unitrel) become
$$
(D - I)\Psi + iL_0\, (D + I) \Psi' = 0\ ,
\eqn\revbcon
$$
or in components,
$$
\eqalign{
\psi(0_+) + L_0 \cot{{\mu_+}\over 2}\, \psi'(0_+) &= 0\ , \cr
\psi(0_-) - L_0 \cot{{\mu_-}\over 2}\, \psi'(0_-) &= 0\ .
}
\eqn\newdiagcon
$$

It is then obvious that the freedom of changing the value
$L_0$ can be absorbed by the corresponding change in the two
parameters $\mu_+$ and $\mu_-$ (by
$\delta\mu_+ = \sin\mu_+ \, \delta L_0 \, / L_0$ and
$\delta\mu_- = \sin\mu_- \, \delta L_0 \, / L_0$
for $L_0 \rightarrow L_0 + \delta L_0$).
This shows that, for describing
distinct boundary conditions, the parameter
$L_0$ does not provide an additional freedom which is
independent of the $U(2)$ parameters
in $U$.

It is interesting to observe that, if both $\mu_+$ and $\mu_-$
are 0 or
$\pi$, a change of $L_0$ does not modify the parameters
$\mu_+$ and
$\mu_-$, that is, a scale change does not affect the
point interaction at all.  These occur at
the values $ (\xi, \alpha_R) = (\frac{\pi}{2}, 0), \, (0, 1) $,
and $(0, -1)$, which correspond exactly to
the scale independent
systems (the continuous family, and the two isolated points,
respectively,
cf.\ Sect.\ 3.1) as expected.

\ve

\secno=0 \appno=2 \meqno=1

\centerline{\bf Appendix B}
\bigskip

In this Appendix we shall present
some basic results concerning the spectral as well as
scattering properties of the
systems with point interaction (see, {\it e.g.},
[\AGHH,\CHughes]).
We first deal with
systems defined on a line, and then
study systems placed in a
box where the entire spectrum becomes discrete.
The latter case is used to demonstrate
the spectral anholonomy and
supersymmetry in Sect.4.

\medskip
\noindent{\bf B.1. System with point interaction on a line}

To begin with, we consider
eigenfunctions with
positive energy $E > 0$.  With
$k = \sqrt{2mE} / \hbar $
the general form of the positive energy eigenfunctions
is given by
    $$
    \varphi_k (x) = \ddef{ A_k^- e^{ikx} + B_k^- e^{-ikx},}
    {x < 0,}{A_k^+ e^{ikx} + B_k^+ e^{-ikx}, }{x > 0,}
    \eqn\eigenpos
    $$
where the coefficients $A_k^\pm$ and $B_k^\pm$ are constants.
This expression, along with the boundary condition
(\unitrel) yields
two linear equations for the coefficients, resulting in
a two-parameter family of solutions $\varphi_k$ to
a given matrix $U$, {\it i.e.},
point interaction.  The spectrum for $E > 0$ is of course
continuous for any $U$, but the characteristics of
the point interaction can still
be seen in the scattering processes.
For this we set $B_k^+ = 0$ and $A_k^- = 1 / \sqrt{2\pi} $
to obtain a plane wave incoming from the left plus
the reflected one in $x < 0$ and the transmitted one in
$x > 0$ by the scattering at $x = 0$,
$$
    \varphi_k^{(l)} (x) = \frac{1}{\sqrt{2\pi}}
\ddef{ e^{ikx} + r^{(l)} e^{-ikx}, }  {x < 0,}
     { t^{(l)} e^{ikx}, \quad \quad } {x > 0.}
    \eqn\psileft
    $$
The reflection and the transmission amplitudes turn
out to be
    $$
    r^{(l)} = \frac{ \alpha \, q + \alpha^* q^{-1} -
(\eta + \eta^*) }{
    \eta \, q + \eta^* q^{-1} - (\alpha + \alpha^*) }\ ,
    \qquad
    t^{(l)} = \frac{ - \beta ( q - q^{-1} ) }{
    \eta \, q + \eta^* q^{-1} - (\alpha + \alpha^*) }\ ,
    \eqn\rltl
    $$
where we have used
$$ \eta = e^{i \xi}\ , \qquad
    q = \frac{ 1 - k L_0 }{ 1 + k L_0 }\ .
    \eqn\etaq
    $$
Similarly, if we set  $A_k^- = 0$
and $B_k^+ = 1 / \sqrt{2\pi} $ we obtain a plane wave
which is incoming from the right and
scattered off at $x = 0$,
    $$
    \varphi_k^{(r)} (x)
= \frac{1}{\sqrt{2\pi}} \ddef{ t^{(r)} e^{-ikx},
    \quad \quad }{x < 0,}{ e^{-ikx} + r^{(r)} e^{ikx}, }{x > 0,}
    \eqn\psiright
    $$
where
    $$
    r^{(r)} = \frac{ \alpha^* q + \alpha \,
    q^{-1} - (\eta + \eta^*) }{
    \eta \, q + \eta^* q^{-1} - (\alpha + \alpha^*) }\ ,
    \qquad
    t^{(r)} = \frac{ \beta^* ( q - q^{-1} ) }{
    \eta \, q + \eta^* q^{-1} - (\alpha + \alpha^*) }\ .
    \eqn\rltl
    $$

The amplitudes $r^{(l)}$, $t^{(l)}$, $r^{(r)}$
and $t^{(r)}$ obey the unitarity conditions,
    $$
    | r^{(l)} |^2 + | t^{(l)} |^2 = 1\ , \qquad
    | r^{(r)} |^2 + | t^{(r)} |^2 = 1\ ,
    \eqn\squaresum
    $$
and
    $$
    {r^{(l)}}^* t^{(r)} + {t^{(l)}}^* r^{(r)} = 0\ .
    \eqn\orthog
    $$
With the aid of these relations and
    $$
    \int_0^\infty dx \, e^{i (k - k') x} = \pi \, \delta (k - k')
    + i \, {\cal P} \frac{1}{k - k'} \qquad ( k, k' > 0 ) \ ,
    \eqn\halfdelta
    $$
we find that $ \varphi_k^{(l)} $ and $ \varphi_k^{(r)} $
are orthonormalized as
    $$
    \int_{-\infty}^\infty dx \, ( \varphi_k^{(l)} )^*
    \varphi_{k'}^{(l)} = \int_{-\infty}^\infty dx \,
    ( \varphi_k^{(r)} )^* \varphi_{k'}^{(r)} = \delta (k - k')\ ,
    \quad \int_{-\infty}^\infty dx\, (\varphi_k^{(l)})^*
    \varphi_{k'}^{(r)} = 0\ .
    \eqn\onrelef
    $$
If we let $k$ take negative values, too, and noting
that $ k \to - k $ implies $ q \to q^{-1} $,
we get further relations
among the reflection and transmission amplitudes as
    $$
    r^{(l)}_{-k} = {r^{(l)}_k}^*, \qquad
    r^{(r)}_{-k} = {r^{(r)}_k}^*, \qquad
    t^{(l)}_{-k} = {t^{(r)}_k}^*, \qquad
    t^{(r)}_{-k} = {t^{(l)}_k}^*\ .
    \eqn\rtkminusk
    $$
A general positive energy eigenfunction (\eigenpos)
is given by the
linear combination of these two solutions,
    $$
    \varphi_k(x) =  C_k^{(l)} \; \varphi_k^{(l)}(x) +
                    C_k^{(r)} \; \varphi_k^{(r)}(x) \ ,
    \eqn\otherphis
    $$
where
the constants $C_k^{(l)}$, $C_k^{(r)}$ are related to
$A_k^\pm$ and
$B_k^\pm$ as $ C_k^{(l)} = \sqrt{2\pi} A_k^-
$ and $ C_k^{(r)} =
\sqrt{2\pi} B_k^+ $.

Turning to the eigenfunctions with
negative energy $E < 0$, we first note
that the general form of a negative energy eigenfunction is
    $$
    \varphi_\kappa (x) = \ddef{ A_\kappa^- e^{\kappa x}
    + B_\kappa^-
    e^{- \kappa x}, }{x < 0,}{
    A_\kappa^+ e^{\kappa x} + B_\kappa^+
    e^{- \kappa x}, }{x > 0,}
    \eqn\eigenneg
    $$
with $\kappa = \sqrt{2m |E|}/ \hbar$.
Since normalizability requires $ A_\kappa^+ = 0 $
and $ B_\kappa^- = 0 $, the boundary condition (\unitrel) gives
    $$
    U \pmatrix{ B^+_\kappa \cr A^-_\kappa } =
    \frac{1 + i \kappa L_0 }{ 1 - i \kappa L_0 }
\pmatrix{ B^+_\kappa \cr A^-_\kappa }.
    \eqn\eigenUBA
    $$
The unit factor on the r.h.s.\ can be written as
$ e^{i \omega_\kappa} $
with $ \omega_\kappa = 2 \arctan \kappa L_0 $.
As $\kappa$ runs in $(0,
\infty)$, $\omega_\kappa$ runs in $ (0, \pi) $.
Thus we find that there can arise maximally two
negative energy bound states under point interactions,
and that there is a one-to-one
correspondence between a bound state
and an eigenvalue
$\lambda$ of the matrix $U$ with
$ \arg \lambda \in (0, \pi) $ (note that $|\lambda| = 1$
since $U$ is unitary).

Under the parametrization
(\stparm), the eigenvalues $\lambda$ of $U$
are determined by the equation,
    $$
    \lambda^2 -
2 \alpha_R e^{i \xi} \lambda + e^{2 i \xi} = 0\ .
    \eqn\chareq
    $$
This has the roots,
    $$
    \lambda_{\pm} = e^{i \xi} \( \alpha_R \pm i
\sqrt{1 - \alpha_R^2} \)\ ,
    \eqn\uxialpha
    $$
or $\lambda_{\pm} = e^{i (\xi \pm \rho)}$ with
$\alpha_R = \cos \rho$ for $ \rho \in [0, \pi] $.  Since
$ \xi - \rho \in [- \pi, \pi) $ and $ \xi + \rho \in [0, 2 \pi)$,
we learn that
there arises a bound state to $\lambda_-$ for $ \xi >
\rho $ and analogously to $\lambda_+$
for $ 0 < \xi + \rho < \pi $.
A doubly degenerate bound state may arise
when $\alpha_R = \pm 1$, namely,
$\rho = 0$ and $\pi$.
The case $\rho = \pi$, however, is not allowed because it
implies $\xi + \rho \geq \pi$, whereas
the case
$\rho = 0$ is allowed for $\xi > 0$.
The latter case in fact belongs to
the self-dual subfamily $\Omega_{SD}$, where now the
characteristic matrix $U$ is given by (\sdsubfamily)
with $\theta \in (0, \pi)$ which is the angle $\xi$ here.
To each eigenvalue $\lambda_\pm$ corresponds the value,
$$
\kappa
= {1\over{iL_0}}\,{{\lambda-1}\over{\lambda+1}}
= {1\over{L_0}}\tan\left({{\xi\pm\rho}\over2}\right)\ .
\eqn\no
$$

The existence of a zero energy $E = 0$ eigenfunction may be
examined by looking at the limit $\kappa \to 0$ of the
negative energy eigenfunctions discussed above.
It is then readily confirmed that
a zero energy state
occurs when there arise roots $\lambda_\pm$ corresponding
to $ \arg \lambda = 0$, that is, at $\xi = \pm \rho$.
In particular, at $\xi = \rho = 0$ there appears a
doubly degenerate state.
Like those appearing for $U$ with negative energy,
this degeneracy
is a consequence of the $su(2)$ spin symmetry
possessed by the interactions of self-dual points
discussed in section 3.  These negative and zero energy
states which are
doubly degenerate give
the ground states for the subfamily
$\Omega_{SD}$, whose energy is given by
    $$
    E_{\rm ground}^{SD} = - \frac{\hbar^2}{2 m L_0^2} \tan^2
    \frac{\theta}{2}\ ,
    \eqn\sdbounden
    $$
for (\sdsubfamily) with $\theta \in [0, \pi)$.

Let us specialize to some of the subfamilies
encountered before and see how the physical quantities
we just obtained look like.  We start with
the scale invariant subfamily $\Omega_W$,
where we have
$\xi = \rho = \pi/2$ and hence no
bound state can arise.
This is in fact expected,
since a bound state energy $E_\kappa$ (or $\kappa$) would
require a length parameter which is lacking
in a scale invariant system.
For the scattering states, one finds that the
reflection and transmission coefficients are
    $$
    r^{(l)} = \alpha_I, \qquad r^{(r)} = - \alpha_I, \qquad
    t^{(l)} = i \beta,  \qquad t^{(r)} = - i \beta^*\ .
\eqn\no
    $$
For the exceptional cases
$U = \pm I$, we obtain
    $$
    r^{(l)} = r^{(r)} = \pm 1, \qquad \quad t^{(l)} = t^{(r)} = 0\ .
\eqn\no
    $$
In all cases, we can observe that the
coefficients are momentum independent, again as a consequence of the
absence of any length parameter.

In the separating (or half-reflection invariant) subfamily
$\Omega_R$, we have the boundary conditions
given by (\vecgenbcwell) which are just
the ones for two `half-line plus infinite wall'
systems.
With
$ L_{\pm} := L_0 \cot(\vartheta_\pm/ 2) $,
the existence of bound states is ensured for
$ 0 < L_+ < \infty $ and $ 0 < L_- < \infty $. If, for
example, $ 0 < L_+ < \infty $ then we find a bound state,
    $$
    \varphi_{\rm bound}^{L_+}(x) = \sqrt{ \frac{2}{L_+} } \;
\Theta(x)
    \; e^{ - x / L_+ }.
\eqn\no
    $$
For the scattering states, on the other hand, we find
    $$
    r^{(l)} = - \frac{1 - i k L_-}{1 + i k L_-}\ , \qquad
    r^{(r)} = - \frac{1 - i k L_+}{1 + i k L_+}\ ,
\eqn\no
    $$
and
    $$
    t^{(l)} = t^{(r)} = 0\ .
\eqn\no
    $$
These results are also in accordance with the ones found for the
`half-line plus infinite wall' systems [\FG].

In the parity invariant subfamily $\Omega_P$ one may put
$ \beta = \beta_I = i \sin \rho $
along with $ \alpha = \alpha_R = \cos \rho $.  One then finds that
bound states exist if $ 0 < \xi \pm \rho \;
(\hbox{mod } 2 \pi) < \pi $, while the scattering states
do not have much
special feature with respect to the generic case,
except that
    $$
    r^{(r)} = r^{(l)} \qquad \hbox{and} \qquad t^{(r)} = t^{(l)}\ .
\eqn\no
    $$
These properties reflect the fact that
the `$ + \leftrightarrow - $' symmetry is just the
`$ r \leftrightarrow  l $' symmetry
of this subfamily.

The subfamily $\Omega_Q$ exhibits
properties similar to $\Omega_P$. With
$ \beta = \beta_R = \sin \rho $, the existence of bound
states is again connected to
the condition, $ 0 < \xi \pm \rho \; (\hbox{mod } 2 \pi) <
\pi $.  The scattering states, while having again the generic
$k$-dependence, possess the amplitudes satisfying
$$
    r^{(r)} = r^{(l)} \qquad \hbox{and}
\qquad t^{(r)} = - t^{(l)}\ .
\eqn\no
$$

The transfer matrix formalism mentioned in Sect.~2 provides
a method which is more appealing
physically than the direct
method we just used.
We demonstrate its
usefulness here
to recover some of the
the scattering
data and the spectral properties first, and
then later use it to study the discrete spectra that appear
under certain Dirichlet and Neumann boundary conditions.
To this end, let us
introduce a vector from
the wavefunction by
$$
\Psi(x)
=  \left( {\matrix{{\varphi (x)}\cr
                  {\varphi' (x)}\cr}
         }
  \right) ,
\eqn\wfv
$$
and define the {\it transfer matrix} $M(x,y)$ by
$$
\Psi(x) = M(x,y) \,\Psi(y) .
\eqn\transfermat
$$
In the absence of interaction, one has
the free transfer matrix,
$$
M_0(x,y) =
\pmatrix{ \cos{k(x-y)}
                & \frac{1}{k}\sin{k(x-y)} \cr
           -k\sin{k(x-y)}
                &  \cos{k(x-y)} \cr} .
\eqn\freetrsf
$$
Consider then the eigenvalue problems
for $M_0(x,y)$ and its conjugate $M_0^\dagger(x,y)$,
$$
M_0(x,y)\,  u_{\pm} = e^{\pm i k(x-y)}\, u_{\pm}\ ,
\qquad
M_0^\dagger(x,y)\,
v_{\pm} = e^{\pm i k(x-y)}\, v_{\pm}\ .
\eqn\treigen
$$
The eigenvectors $u_{\pm}$ and $v_{\pm}$ are
given by
$$
u_{\pm}
= \frac{1}{\sqrt{2}}
  \left( {\matrix
           {{1}\cr
            {\pm i k}\cr}
         }
  \right) ,
\qquad
v_{\pm}
= \frac{1}{\sqrt{2}}
  \left( {\matrix
           {{1}\cr
            {\mp 1/i k}\cr}
         }
  \right) ,
\eqn\eigvect
$$
which satisfy the bi-orthogonal relations,
$$
v_{\pm}^\dagger u_\pm = 1 ,
\qquad
v_{\mp}^\dagger u_\pm = 0 .
\eqn\biorth
$$
The advantage of using the transfer matrix is that
the boundary condition (\trcon) characteristic to
the point interaction --- which is a local
description of (\unitrel) --- provides precisely the
connection condition for the transfer matrix
at $x=0_-$ and $0_+$ as
$M(0_+,0_-) = \Lambda$ or
$$
\Psi(0_+) = \Lambda \,\Psi(0_-)\ ,
\eqn\inftrsf
$$
which is nothing but the condition (\trcon).

By means of the transfer matrix, the
scattering process induced by
an incoming plane wave from the right, for instance,
can easily be determined as follows.   First, the
vector (\wfv) corresponding to the wavefunction (\psiright)
is given by
$$
    \Psi_k^{(r)} (x)
= \frac{1}{\sqrt{2\pi}} \ddef{ t^{(r)} e^{-ikx}u_-,
    \quad \qquad }{x < 0,}{ e^{-ikx}u_-
+ r^{(r)} e^{ikx}u_+, }{x > 0,}
    \eqn\no
    $$
At the point singularity, the vector must satisfy
(\inftrsf) which reads
$$
u_- + r^{(r)} u_+ = t^{(r)} \Lambda u_-\ .
\eqn\no
$$
{}From this, with the help of bi-orthogonality, we obtain
immediately the amplitudes,
$$
t^{(r)} = {{1}\over{v_-^\dagger \Lambda u_-}}\ , \qquad
r^{(r)} = {{v_+^\dagger \Lambda u_-}\over{v_-^\dagger \Lambda u_-}}\ ,
\eqn\amptf
$$
where we have
$$
\eqalign{
v_-^\dagger \Lambda u_-
&= {{\Lambda_{11}+\Lambda_{22}}\over{2}}
-i{{\Lambda_{12}k-\Lambda_{21}/k}\over{2}}\ ,
\cr
v_+^\dagger \Lambda u_-
&= {{\Lambda_{11}-\Lambda_{22}}\over{2}}
-i{{\Lambda_{12}k+\Lambda_{21}/k}\over{2}}\ ,
}
\eqn\no
$$
in terms of the components of $\Lambda$.  It can be
readily confirmed using (\matl) that
these amplitudes in (\amptf) agree with the ones
in (\rltl) obtained under the global description.
The negative energy bound states can also be
found from the amplitude $t^{(r)}$ (or $r^{(r)}$)
by looking at the poles on the
imaginary axis in the complex $k$-plane.

It is instructive to look at the scattering amplitudes
for the two limiting cases of delta and epsilon function
potentials.  For delta function potential, $\Lambda_{21}=0$,
the wave function is symmetric, {\it i.e.},
$$
\Psi(0_+) =
  \left( {\matrix
           {{\varphi_+(0_+)}\cr
            {\varphi'_+(0_+)}\cr}
         }
  \right) ,
\qquad
\Psi(0_-) =
  \left( {\matrix
           {{\varphi_+(0_+)}\cr
            {-\varphi'_+(0_+)}\cr}
         }
  \right) .
\eqn\no
$$
{}From this and eqs.(\pbcwell) and (\chiralcc), one has
$$
\Lambda =
  \left( {\matrix
           {{1}&{0}\cr
            {-2g_+/L_0}&{1}\cr}
         }
  \right) ,
\eqn\no
$$
which results in
$$
t^{(r)} = {{1}\over{1-ig_+/(kL_0)}}\ , \qquad
r^{(r)} = {{1+ig_+/(kL_0)}\over{1-ig_+/(kL_0)}}\ .
\eqn\no
$$
Thus we have $|t^{(r)}|^2 = 0$ at $k=0$ and
$|t^{(r)}|^2\to 1$ at $k\to \infty$. In other words,
the delta function potential works as a high-pass
(or low-cut) filter
for the incoming quantum wave.
Similarly, for epsilon function potential,
$\Lambda_{12}=0$, whose wave functions
are antisymmetric,
$$
\Psi(0_+) =
  \left( {\matrix
           {{\varphi_-(0_+)}\cr
            {\varphi'_-(0_+)}\cr}
         }
  \right) ,
\qquad
\Psi(0_-) =
  \left( {\matrix
           {-{\varphi_-(0_+)}\cr
            {\varphi'_-(0_+)}\cr}
         }
  \right) ,
\eqn\no
$$
and therefore
$$
\Lambda =
  \left( {\matrix
           {{1}&{-2g_-L_0}\cr
            {0}&{1}\cr}
         }
  \right) ,
\eqn\no
$$
one obtains
$$
t^{(r)} = {{1}\over{1+ig_-kL_0}}\ , \qquad
r^{(r)} = {{1-ig_- kL_0}\over{1+ig_- kL_0}}\ ,
\eqn\no
$$
thus $|t^{(r)}|^2 = 1$ at $k=0$ and
$|t^{(r)}|^2\to 0$ at $k\to \infty$.  One thus sees, for
instance,
that the $\varepsilon$-function potential works as a
low-pass (or high-cut) filter.
In the general case (\amptf), one has a quantum filter that
passes only certain range of momentum value cutting out
both high and low frequency components
from the transmitted waves.

\medskip
\noindent{\bf B.2. System with point interaction in a box}

So far our system has been that of a particle
moving freely on a line $\R$ under a point singularity
at $x = 0$.  However, even if we put the particle
in a box given, say,
by the interval $-l_- \le x \le l_+$,
the effect of
the singularity at $x = 0$ remains the same and
can be characterized by the same condition (\unitrel).
The extra requirement we need to take into account
is the boundary conditions at the edges
$x=l_+$ and $x=-l_-$, and for this we shall consider
some typical cases below.

We begin by the simple
case $l_+ = l_- = l$ with a periodic or
antiperiodic boundary condition,
$\Psi(l) = \pm \Psi(-l)$, that is,
$$
\Psi(0_-) =  \pm M_0(2l)\, \Psi(0_+)\ .
\eqn\no
$$
Combining this with the
connection condition (\inftrsf) at the singularity,
we obtain the eigenvalue problem,
$$
M_0(-2l)\,\Psi(0_-) = \pm \Lambda\, \Psi(0_-)\ .
\eqn\no
$$
Energy eigenstates arise at the roots for $k$
in the eigenvalue equation,
$$
\det[\Lambda \mp M_0(-2l)] = 0\ ,
\eqn\no
$$
which is
$$
2\sin\xi \cos{2kl}
+\left[
   \cos\xi \left( kL_0+\frac{1}{kL_0} \right)
   +\alpha_R \left( kL_0-\frac{1}{kL_0} \right)
 \right] \sin{2kl}
\mp 2 \beta_I
= 0 .
\eqn\no
$$
The periodic case is equivalent to the system
of a circle of length $2l$, for which
the spectral classification has been
discussed in various subfamilies in [\FT].

Next we consider two cases of an ideal
box where \lq infinite' walls
are placed at the edges so that no
probability current can leak from the box.
One of them is given by
the Dirichlet
boundary condition
$\varphi(l_+) = \varphi(-l_-) = 0$
at the edges.
In terms of the transfer matrix, we then have
$$
M_0(l_+) \,\Lambda \,M_0(l_-)
\left( {\matrix{{0}\cr
                {\varphi' (l_-)}\cr}
       } \right)
=
\left( {\matrix{{0}\cr
                {\varphi' (l_+)}\cr}
       } \right),
\eqn\no
$$
from which we obtain the eigenvalue equation,
$$
\left( M_0(l_+) \,\Lambda\, M_0(l_-)
\right)_{12} = 0\ .
\eqn\no
$$
Similarly, for the Neumann boundary condition
$\varphi'(l_+) = \varphi'(-l_-) = 0$
we find
$$
M_0(l_+)\, \Lambda\, M_0(l_-)
\left( {\matrix{{\varphi(l_-)}\cr
                {0}\cr}
       } \right)
=
\left( {\matrix{{\varphi(l_+)}\cr
                {0}\cr}
       } \right)\ ,
\eqn\no
$$
which leads to
$$
\left( M_0(l_+)\, \Lambda\,
M_0(l_-) \right)_{21} = 0\ .
\eqn\no
$$
In each case, the spectrum is determined by
the eigenvalue equation.  Explicitly, for the case of
the Dirichlet boundary condition, the equation
leads to
$$
\eqalign{
&(kL_0)^2 \cot{kl_+} \cot{kl_-} (\cos\xi+\alpha_R)
\cr
&- kL_0 \left[\cot{kl_+} (\sin\xi-\alpha_I)
            +\cot{kl_-} (\sin\xi+\alpha_I)
       \right]
-(\cos\xi-\alpha_R)
= 0\ ,
}
\eqn\no
$$
whereas for the case of the Neumann boundary
condition, it yields
$$
\eqalign{
&(kL_0)^2 \tan{kl_+} \tan{kl_-} (\cos\xi+\alpha_R)
\cr
&+ kL_0 \left[\tan{kl_+} (\sin\xi+\alpha_I)
            +\tan{kl_-} (\sin\xi-\alpha_I)
       \right]
-(\cos\xi-\alpha_R)
= 0\ .
}
\eqn\no
$$
In particular for $l_+ = l_- = l$, we have
$$
kL_0 \cot{kl} = { { \sin\xi \pm \sqrt{1 - \alpha_R^2} }
                \over { \cos\xi + \alpha_R } }
\eqn\spcondir
$$
for the Dirichlet case, and
$$
kL_0 \tan{kl} = { { - \sin\xi \pm \sqrt{1 - \alpha_R^2} }
                \over { \cos\xi + \alpha_R} }
\eqn\no
$$
for the Neumann case, respectively.  In particular, for
$\alpha_R = 0$, $\xi=\pi/2$, {\it i.e.,}
in the scale invariant subfamily $\Omega_W$,
one simply has
$$
\cos{k l} = 0\ ,
\eqn\no
$$
for the Dirichlet case, and
$$
\sin{k l} = 0\ ,
\eqn\no
$$
for the Neumann case, which shows that the scale invariant
sphere $\Omega_W \simeq S^2$ in $\Omega$ is isospectral.


\ve
\baselineskip= 15.5pt plus 1pt minus 1pt
\parskip=5pt plus 1pt minus 1pt
\tolerance 8000
\vfill\eject

  \vfill\eject\immediate\closeout\reffile
  \centerline{{\bf References}}\bigskip\frenchspacing%
  \input refs.tmp\vfill\eject\nonfrenchspacing

\bye